\documentclass[a4paper,fleqn,usenatbib,useAMS]{mnras}

\usepackage{graphicx}
\usepackage{amsmath}	
\usepackage{amssymb}	
\usepackage{multicol}        
\usepackage{bm}		
\usepackage{pdflscape}	
\usepackage[encapsulated]{CJK}
\usepackage{ucs}
\usepackage[utf8x]{inputenc}
\usepackage{xcolor}

\usepackage[T1]{fontenc}
\usepackage{ae,aecompl,ulem}

\usepackage{txfonts}

\title[Dust in NGC\,6888]{Unveiling the stellar origin of the Wolf-Rayet nebula NGC\,6888 through infrared observations}
\author[G.\ Rubio et al.]{G.\,Rubio$^{1,2}$\thanks{E-mail: grubio@idec.edu.mx}, J.A.\,Toal\'{a}$^{3}$\thanks{E-mail: j.toala@irya.unam.mx}, P.\,Jim\'{e}nez-Hern\'{a}ndez$^{3}$, G.\,Ramos-Larios$^{1,2}$, M.A.\,Guerrero$^{4}$,      
\newauthor{V. M. A.\,Gómez-González$^{3}$, E.\,Santamar\'\i a$^{1,2}$ and J.A.\,Quino-Mendoza$^{1,2}$}\\
$^1$CUCEI, Universidad de Guadalajara, Blvd. Marcelino Garc\'\i a Barrag\'an 1421, 44430, Guadalajara, Jalisco, Mexico \\
$^2$Instituto de Astronom\'\i a y Meteorolog\'\i a, Dpto.\ de F\'\i sica,
CUCEI, Av.\ Vallarta 2602, 44130, Guadalajara, Jalisco, Mexico\\
$^3$Instituto de Radioastronom\'{i}a y Astrof\'{i}sica (IRyA), UNAM Campus Morelia,
Apartado postal 3-72, 58090 Morelia, Michoac\'{a}n, Mexico\\
$^4$Instituto de Astrof\'\i sica de Andaluc\'\i a, IAA-CSIC, Glorieta de la Astronom\'\i a s/n, 18008, Granada, Spain\\
}

\pubyear{2020}

\begin{document}
\label{firstpage}
\pagerange{\pageref{firstpage}--\pageref{lastpage}}
\maketitle

\begin{abstract}
We present a comprehensive infrared (IR) study of the iconic Wolf-Rayet
(WR) wind-blown bubble NGC\,6888 around WR\,136. 
We use {\it Wide-field Infrared Survey Explorer} ({\it WISE}), 
{\it Spitzer} IRAC and MIPS and {\it Herschel} PACS
IR images to produce a sharp 
view of the
distribution of dust around WR\,136. 
We complement these IR photometric observations
with {\it Spitzer} IRS spectra in the 5--38~$\mu$m wavelength range.
The unprecedented high-resolution IR images allowed us to produce a clean 
spectral energy distribution, free of contamination from material along the 
line of sight, to model the properties of the dust in NGC\,6888.  
We use the spectral synthesis code Cloudy to produce a model for NGC\,6888 
that consistently reproduces its optical and IR properties. Our best model
requires a double distribution with the inner shell composed only of gas, whilst the outer shell requires a mix of gas and dust.  
The dust consists of two populations of grain sizes, one with small sized 
grains $a_\mathrm{small}$=[0.002--0.008]~$\mu$m and another one with large 
sized grains $a_\mathrm{big}$=[0.05--0.5]~$\mu$m.
The population of big grains is similar to that reported for other red 
supergiants stars and dominates the total dust mass, which leads us to 
suggest that the current mass of NGC\,6888 is purely due to material 
ejected from WR\,136, with a negligible contribution of swept up 
interstellar medium.
The total mass of this model is 25.5$^{+4.7}_{-2.8}$~M$_{\odot}$, a
dust mass of $M_\mathrm{dust}=$0.14$^{+0.03}_{-0.01}$~M$_{\odot}$, for 
a dust-to-gas ratio of $5.6\times10^{-3}$.
Accordingly, we suggest that the initial stellar mass of WR\,136 was 
$\lesssim$50~M$_{\odot}$,
consistent with current single stellar evolution models.

\end{abstract}

\begin{keywords}
stars: evolution --- stars: individual: NGC\,6888, WR\,136 -- 
stars: winds, outflows --- infrared: ISM --- stars: Wolf-Rayet
\end{keywords}



\section{INTRODUCTION}
\label{sec:intro}

Wolf-Rayet (WR) stars are formed as the result of the evolution of
O-type stars with intial masses
$>20$~M$_{\odot}$ \citep[][]{Ek2012}.
The single stellar evolutionary scenario suggests that,  
before entering the WR phase, these stars become red supergiant (RSG) or luminous blue
variable (LBV) stars whose slow winds ($v_{\infty}\approx30-200$~km~s$^{-1}$) 
strip them from their hydrogen-rich outer layers producing
dense structures around them \citep[e.g.,][]{Cox2012,Morris2017}. 
Furthermore, observational evidence suggests that the majority of
massive stars are born in binary systems \citep{Sana2012}, which
affect their evolution and, consequently, the production of WR
stars \citep{Eldridge2017,Mason2009}.
Mass transfer between the binary components or the ejection of a
common envelope might also strip the hydrogen-rich envelope of one
of the components creating a WR star 
\citep[see, e.g.,][and references therein]{Gotberg2018,Ivanova2011}. 
Nevertheless, WR stars exhibit strong fast winds
\citep[$v_{\infty}\approx$1500~km~s$^{-1}$,
  $\dot{M}$$\approx$10$^{-5}$~M$_{\odot}$~yr$^{-1}$;][]{Hamann2006} that sweep
and compress the previously ejected material. 
This interaction produces a wind-blown bubble around the WR star 
and, at the same time, the strong UV flux from the progenitor star ionizes 
the material \citep[e.g.,][]{GarciaSegura1995}. 
The combination of all these effects produces WR nebulae \citep[see][]{Chu1981}.

WR nebulae have radii as large as $\sim$10~pc
\citep[e.g.,][]{Toala2012,Toala2017} and in some cases
exhibit multiple shells due to the eruptive mass ejection of the progenitor star
\citep{Marston1995a,Marston1999}. 
This demonstrate the powerful energy injection of 
very massive stars leveraging their role as one of the main actors shaping and enriching the 
interstellar medium (ISM). 
Spectroscopic studies of WR nebulae can unveil the production 
of heavy elements in their interiors, which can be used to test stellar
evolution models and the chemical gradients in their host galaxies
\citep[][]{Esteban2016,MD2020}.
However, these structures are short-lived and therefore 
seldomly detected. WR nebulae will 
experience hydrodynamical instabilities causing them 
to break and to dissipate into the ISM within a few 
times 10$^{4}$~yr \citep[][]{Freyer2006,Toala2011}. 
That is, not many WR stars exhibit associated
nebulae \citep{Gruendl2000,Stock2010}, thus, producing detailed studies of known
WR nebulae can bring us closer to understanding the 
violent impact of massive stars in the ISM.

Massive stars have been appointed to be 
laboratories for studying dust formation, processing and evolution 
\citep{Watcher2010,Gv2010,Ver2009}. 
During the RSG and LBV phases these stars reduce their effective 
temperature, $T_{\mathrm{eff}}\lesssim10^{4}$~K, allowing the formation 
of dust in their surroundings.  
The details of these processes depend on whether the star
evolved through a RSG or LBV phase. 
RSG stars, with their enhanced mass-loss rates 
($\dot{M} \lesssim 10^{-4}$~M$_{\odot}$~yr$^{-1}$),
may form dust-rich shells with dust sizes as large as a few times
0.1~$\mu$m \citep[see][and references therein]{Scicluna2015}. 
Meanwhile, the eruptive nature of the high mass-loss rate of LBV stars 
\citep[$\dot{M} \gtrsim 10^{-3}$~M$_\odot$~yr$^{-1}$;][]{Weis2001} may 
cause the dust to be shielded from stellar radiation, allowing it to 
grow to sizes as large as 1~$\mu$m \citep[][]{Kochanek2011} as reported 
for the iconic LBV star $\eta$~Car \citep{Morris2017}.
Dust production can also occur at the interacting wind surface of massive binaries.
In particular, carbon-rich WR stars (WC) with OB companions have been found to exhibit
dust-rich pinwheel nebulae \citep[e.g.,][]{Tuthil1999,Marchenko2002}. Furthermore,
it has been suggested that dust can also form in the ejecta of a common envelope phase 
around a binary system \citep{Lu2013}. 

Dust has been found spatially associated with the 
brightest optical regions of WR nebulae. 
Early infrared (IR) studies of WR nebulae showed that dust is thermally-heated by
the strong ionizing flux from their progenitor stars \citep{vanBuren1988}.
Larger cavities harbouring WR nebulae
have also been detected through IR and radio observations and these
very likely correspond to the previous bubble carved
by the feedback of the progenitor star during the main 
sequence phase \citep{Marston1995b,Marston1996,Arnal1996,Cappa1996}. 
\citet{Marston1991} presented the analysis of 
far-IR {\it IRAS} observations of the WR nebulae 
NGC\,2359, RCW\,58 and NGC\,6888 and
suggested that their dust masses were in the range of 0.25--1.3~$M_{\odot}$. 
Taking into account the typically-adopted dust-to-gas ratio of 0.01, 
the total nebular masses of these WR nebulae
were too large to have a stellar origin.  
Accordingly, \citet{Marston1991} attributed them to swept up material 
from the ISM. 
A subsequent study presented by \citet{Mathis1992} demonstrated that 
these mass estimates can change dramatically if the 
effects of the radiation field from the star are accounted for.

\begin{figure*}
\begin{center}
\includegraphics[width=0.9\linewidth]{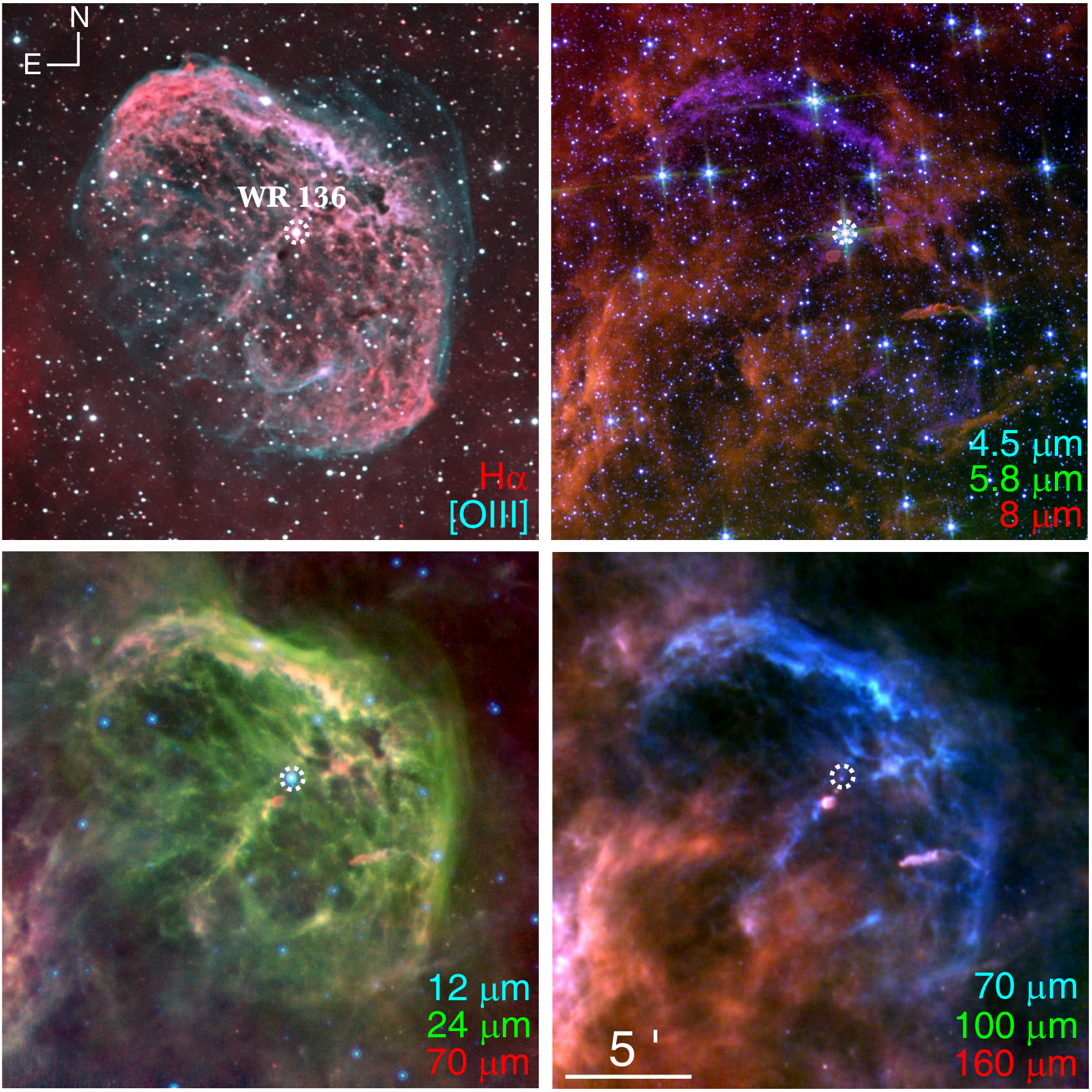}
\caption{
Optical and IR views of the WR nebula NGC\,6888 showing different morphological 
structures and the spatial distribution of dust. 
{\it Top left}: Colour-composite optical image. 
{\it Top right}: Composition of {\it Spitzer} IRAC images. 
{\it Bottom left}: Composition of different mid-IR instruments ({\it WISE}, 
{\it Spitzer} MIPS and {\it Herschel} PACS). 
{\it Bottom right}: Composition of {\it Herschel} PACS images. 
The position of WR\,136 is shown with a dashed-line circle. 
All panels have the same field of view.
}
\label{fig:mosaic}
\end{center}
\end{figure*}

\begin{table*}
\begin{center}
\caption{Infrared observations of NGC 6888 used in this work.}
\setlength{\tabcolsep}{0.8\tabcolsep}    
\begin{tabular}{lcccccl}
\hline
Telescope & Instrument & Band                   & Obs.\,date    & Obs.\,ID. & Processing level & PI \\
          &            & ($\mu$m)               &  (yyyy-mm-dd) &           &                  &    \\
\hline
{\it Spitzer} & IRAC   &3.6, 4.5, 5.8, 8.0      & 2005-10-21    & 20726     & Level 2          & J. Hester \\
              & MIPS   & 24                   & 2005-09-23    & 20726     & Level 2          & J. Hester \\
              & IRS    & SL (5--15), LL (14--39) & 2006-11-09    & 30544     & Level 2         & P. Morris \\
\hline
{\it Herschel} & PACS  & 70, 160                & 2010-12-16    & 1342212042 &   Level 3        & M. Groenewegen  \\
               &       & 100, 160               & 2010-12-16    & 1342212040 &   Level 3        & M. Groenewegen  \\    
\hline
{\it WISE} &           & 12                     & 2010-05-14    & 3031p378\_ac51 & Level 3  &            \\
\hline
\end{tabular}
\label{tab:table1}
\end{center}
\end{table*}

The advent of a new generation of IR satellites provided us with 
the opportunity of spatially resolve structures in WR nebulae, 
improving our understanding of dust processing 
in nebulae around massive stars \citep[see][]{Cicho2020}. 
Furthermore, {\it Spitzer} and {\it WISE} have been used to unveil a new 
population of obscured nebulae around LBV and WR stars 
\citep{Gv2010,Toala2015}. More recently, {\it Herschel} observations of M\,1-67 
around WR\,124 were presented by \citet{VN2016}.
Their radiative transfer model predicted a total nebular 
mass of 22~$M_{\odot}$ with a dust mass of 0.26~$M_{\odot}$. 
These authors predicted dust particles 
with sizes between 2 and 10~$\mu$m in M\,1-67, but such large 
dust particles have never been reported in 
circumstellar nebulae.

We have started a series of works to analyze publicly available IR
observations of WR nebulae using the photoionization
code {\sc Cloudy} \citep[][]{Ferland2017} in conjunction with detailed
synthetic WR spectra obtained from the state-of-the-art Potsdam Wolf-Rayet Models (PoWR) models
\citep[][]{Hamann2004}\footnote{\url{http://www.astro.physik.uni-potsdam.de/PoWR}}. 
Our most recent work on the WR nebula M\,1-67 showed that there is no need for 
dust with sizes larger than 1~$\mu$m to reproduce the nebular and dust properties 
\citep[][hereinafter Paper~I]{Palmira2020}. 
By combining modern tools we were able to conclude that the origin of
M\,1-67 might have been due to a common envelope scenario, which
makes this WR nebula the first evidence of such process in massive
stars.

In this paper we present our analysis of the WR nebula NGC\,6888
(a.k.a. the Crescent Nebula; see Fig.~\ref{fig:mosaic} top left) 
around WR\,136
(a.k.a. HD\,192163), one of the most studied galactic WR
nebulae. H$\alpha$ narrow-band emission images show a filamentary and
clumpy structure \citep[][]{Stock2010}, which very likely formed as a result of
hydrodynamical instabilities generated through the wind-wind interacting scenario
\citep[e.g.,][]{GS1996,Toala2011}. 
An outer {\it skin} of [O~{\sc iii}] emission encompassing the clumpy structure 
results from the shock of the expanding RSG material pushed by the WR wind
\citep[][]{Gruendl2000,moore2000}.  
This [O~{\sc iii}] shell also confines the adiabatically-shocked, X-ray-emitting hot 
bubble \citep[see][and references therein]{Toala2016,Toala2014,Wrigge2005}.

Several studies have addressed the physical properties of NGC\,6888 and in 
general all these studies result in similar abundances 
\citep[see][and references therein]{Esteban2016,fernandez2012,reyes2015,Stock2014}.
\citet[][]{fernandez2012} suggested that NGC\,6888 is mainly composed of 
three structures: 
i) an ellipsoidal inner broken structure formed by shocked shells from the WR 
and RSG stages, 
ii) an outer spherical shell presumably formed by the breaking of the main 
sequence (MS) bubble, and 
iii) a faint structure around the nebula created by the interaction of the MS 
winds and the local ISM. 
Using the ionization code {\sc Cloudy}, \citet[][]{reyes2015} concluded that 
NGC\,6888 is chemically homogeneous and uniformly filled with low density 
material, 1~cm$^{-3}$, in stark contrast to the obvious different morphological 
components and previous density estimates.

Here we present a complete analysis of the IR emission of
NGC\,6888. We use images and spectra from {\it WISE}, {\it Spitzer} 
and {\it Herschel} that cover the
3--160~$\mu$m wavelength range. The images are used to study the
distribution of dust in NGC\,6888 in comparison with optical images. 
The spectral energy distribution (SED) in the IR, in combination with publicly 
available estimates of the abundances and nebular properties, has been 
used to model the properties of the dust and ionized components of NGC\,6888 
with the ionization code {\sc Cloudy}. 
Our paper is organised as follows. The
observations are presented in Section~2. 
The analyses of the images, the IR photometry, and the IR spectra are presented
in Sections~3, 4 and 5, respectively. 
Our models of NGC\,6888 are presented in Section~6.
Finally, the discussion and conclusions are presented in Sections~7 and 8.

\section{Observations}
\label{sec:observations}
\begin{figure*}
\begin{center}
  \includegraphics[width=0.96\linewidth]{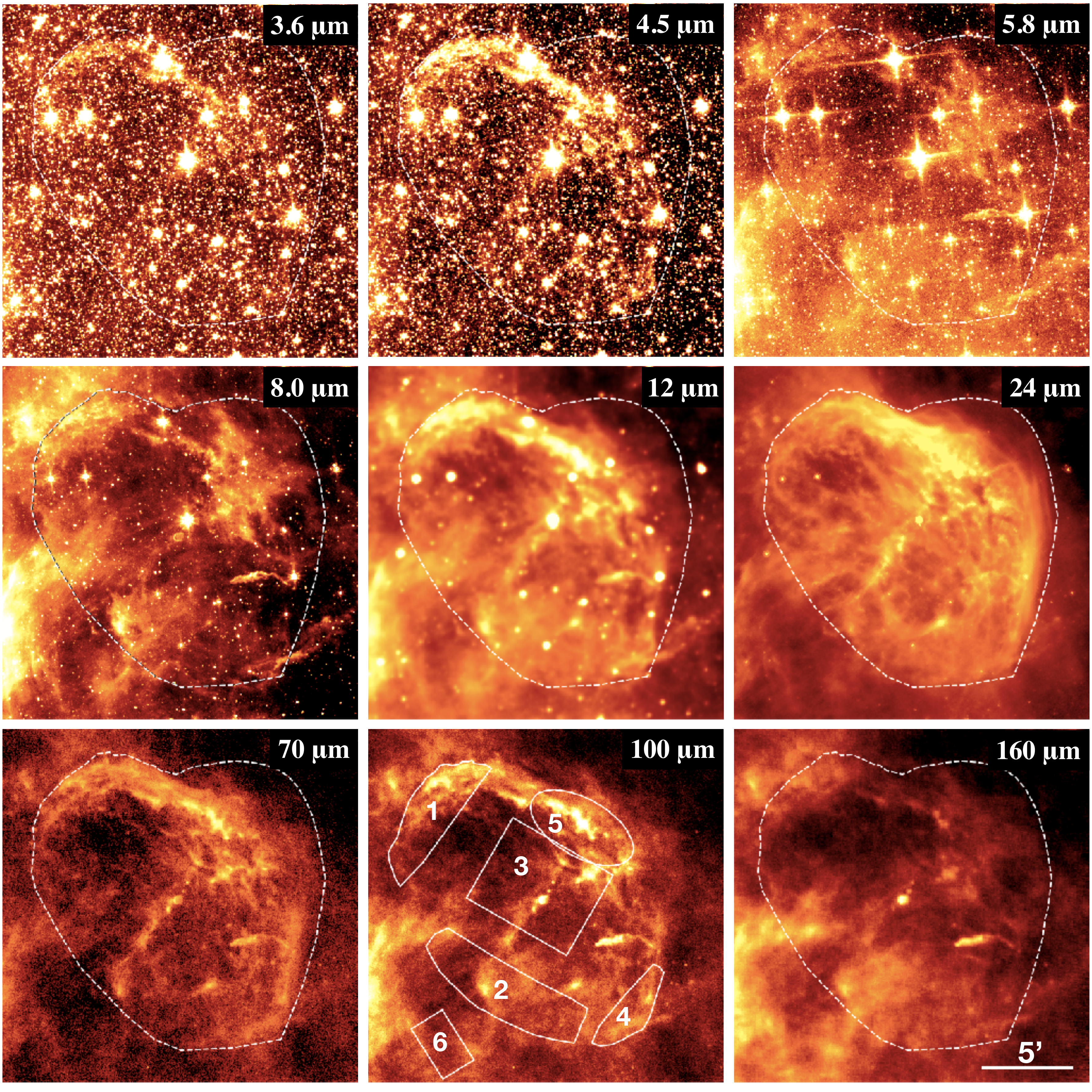}
\caption{IR views of the WR nebula NGC\,6888. Each panel 
correspond to different IR filters (see Table~1). 
The dashed-line 
region delimits the optical [O\,{\sc iii}] emission 
detected in the optical (see Fig.~\ref{fig:mosaic}). Regions
labelled from 1 to 6 were used to derive the photometric measurements
presented in Figure~\ref{fig:sed}. All panels have the same FoV
as those in Figure~\ref{fig:mosaic}. North is up, east to the left.}
\label{fig:panelIR}
\end{center}
\end{figure*}

\subsection{IR images and spectra}

All the IR observations used in the present work have been obtained from
the NASA/IPAC Infrared Science
Archive\footnote{\url{https://irsa.ipac.caltech.edu/frontpage/}}. 
Details of the observations such as the telescope, instrument, 
spectral band, observation date and principal investigator (PI)
are listed in Table~\ref{tab:table1}. Archival IR images of NGC\,6888 were obtained from
{\it WISE}, {\it Spitzer} IRAC \& MIPS, and {\it Herschel} PACS 
observations\footnote{The {\it Herschel} PACS 
observations were obtained in the framework of the Mass-loss of Evolved StarS 
(MESS) project \citep{Groe2011}.}. 
All available IR images are presented in Figure~\ref{fig:panelIR}.

The available {\it Spitzer} Infrared Spectrograph (IRS) 
observations of NGC\,6888 were 
obtained in {\it stare} mode \citep[][]{houck2004}. 

These spectra were obtained with the low- and high-resolution modules, but 
unfortunately the high-resolution observations do not have suitable regions 
for background-subtraction. 
Thus, we inspected these spectra, but no science was obtained from them.
The low-resolution spectra correspond to the Short-Low (SL) and Long-Low 
(LL) modules covering the 5--38~$\mu$m wavelength range. 
The positions of the low-resolution slits are shown in 
Figure~\ref{fig:low_res_positions}.

The {\it Spitzer} IRS observations were analysed with 
the CUbe Builder for IRS Spectra Maps 
\citep[{\sc cubism};][]{smith2007} software. 
{\sc cubism} tasks are used to combine spectra from 
the same position and to produce one-dimensional spectra. 
We followed the standard reduction processes, which 
include background subtraction, characterization
of noise in the data and bad pixel removal.

\subsection{Optical imaging (OARP)}

Optical narrow-band images of NGC\,6888 were obtained at the Observatorio 
Astron\'omico Rob\'otico Primavera (OARP) operated by the University of 
Guadalajara (Jalisco, Mexico). The facility is used by undergraduate,
graduate, and doctoral programs in physics, astronomy and astrophysics.
Recently established, was strategically located just over 20
miles west of Guadalajara, where it was free from light pollution
and clearer night sky.
The observatory utilizes a 0.32m f/8 CDK (Corrected Dall-Kirkham) telescope 
equipped with a SBIG STL-6303E/LE commercial camera, which uses a Kodak
Enhanced KAF-6303E imaging sensor. 
This sensor consists of a 3072$\times$2048 pixels array with a size of 
9$\times$9 microns each.  
In combination with the telescope, it provides a field of view (FoV) of 
$37'\times25'$ and a plate scale of 0$\farcs$73~pixel$^{-1}$.

Images of NGC\,6888 were obtained on 2019 March 9 and 30 for total exposure 
times of 2~h in the [O~{\sc iii}] filter and 1.5~h in the H$\alpha$ filter.  
These filters have central wavelengths and bandwidths of 
$\lambda_\mathrm{c}$ = 6563~\AA\ and $\Delta\lambda$=30~\AA, and 
$\lambda_\mathrm{c}$ = 5007~\AA\ and $\Delta\lambda$=30~\AA, respectively. 
The detector was set with a $2\times2$ binning. 
The data were bias-subtracted and flat-field corrected following standard Image Reduction 
and Analysis Facility ({\sc iraf}) routines \citep[][]{Tody1993}. 
A colour-composite optical image of NGC\,6888 is presented in the top-left panel 
of Figure~\ref{fig:mosaic}.

\section{Dust distribution}

\begin{figure}
\begin{center}
\includegraphics[width=\linewidth]{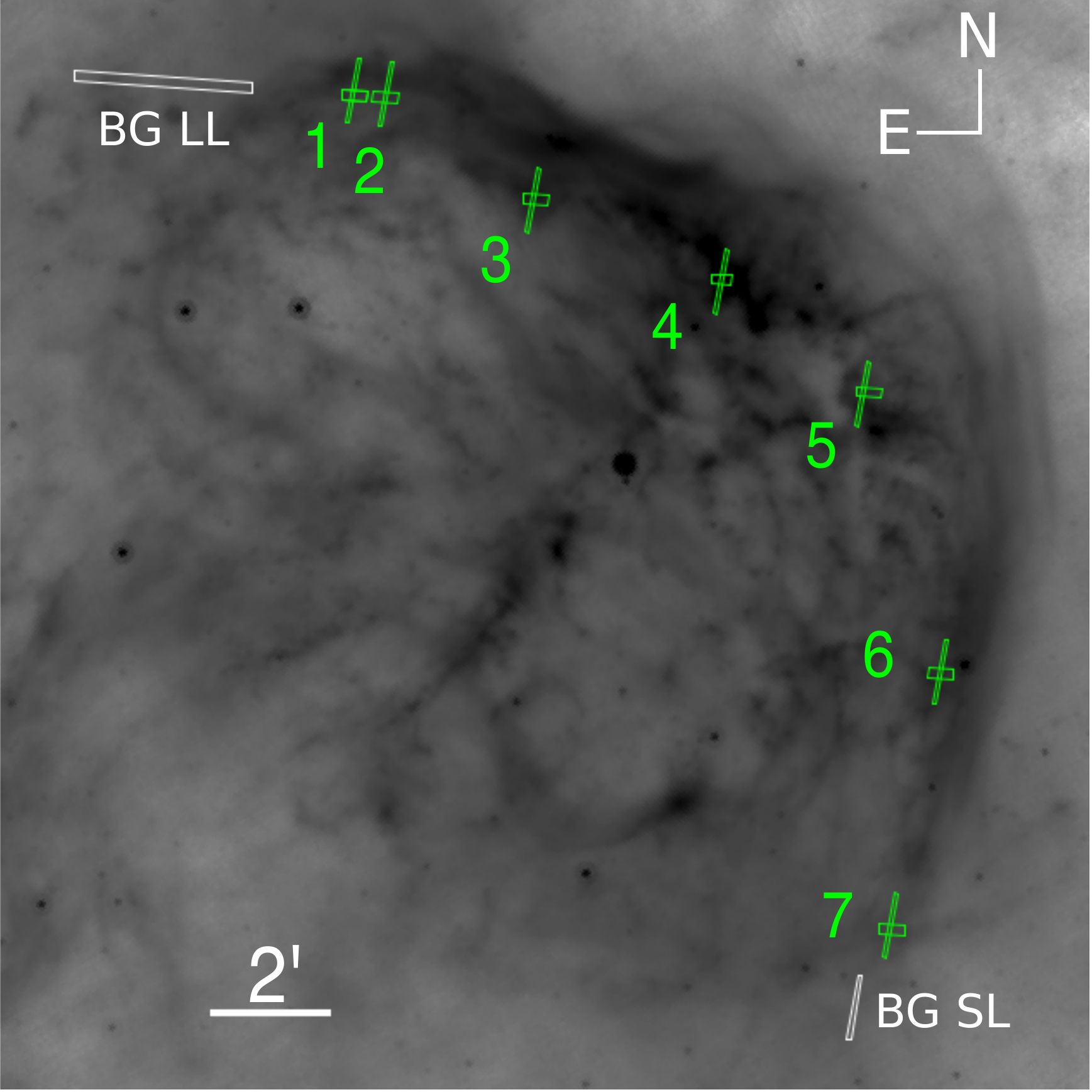}
\caption{{\it Spitzer} MIPS 24~$\mu$m image of NGC\,6888 
showing the position of the SL and LL slits (green). 
The one oriented north to south correspond to SL, the one oriented east to west is LL.
The regions used to extract the background are 
  shown with white rectangles.}
\label{fig:low_res_positions}
\end{center}
\end{figure}

Figure~\ref{fig:panelIR} presents the different IR images of NGC\,6888 
in the 3--160~$\mu$m wavelength range. 
All {\it Spitzer} IRAC images exhibit a large number of point sources in 
the vicinity of NGC\,6888.  
The IRAC 3.6 and 4.5~$\mu$m images show a marginal detection 
coincident with the brightest optical regions of NGC\,6888. 
Some emission is detected from the NE cap and in a lesser 
extent from the SW cap. No significant emission from the inner 
region is detected. 
The other two IRAC images, 5.8~$\mu$m and 8.0~$\mu$m, trace mostly clumps 
and filaments of the cold surrounding ISM. 

The {\it Spitzer} MIPS image at 24~$\mu$m traces the continuum emission from 
dust present in the WR nebula \citep{Toala2015}. 
 
The {\it Herschel} PACS 70~$\mu$m image is very similar to that of the 
{\it Spitzer} MIPS, but the PACS 160~$\mu$m image rather traces the 
emission from the cold ISM.  
Meanwhile, the emission in the {\it WISE} 12~$\mu$m and {\it Herschel} PACS 
100~$\mu$m images map the contribution from both the nebula and the ISM.

In general, most images presented in Figure~\ref{fig:panelIR} 
trace the morphological features exhibited by the optical 
image of NGC\,6888: the NE and SW caps and the NW blowout. 
The dark clump and filament described by \citet{fernandez2012} are clearly 
visible in the IRAC 5.8~$\mu$m and 8.0~$\mu$m images and in the {\it Herschel} 
images, they form part of the cold ISM. 
For a direct comparison with the optical image, we created three colour-composite 
IR pictures of NGC\,6888 using 
i) the 4.5, 5.8 and 8~$\mu$m {\it Spitzer} IRAC images, 
ii) the 12, 24 and 70~$\mu$m mid-IR images, and 
iii) the 70, 100 and 160~$\mu$m {\it Herschel} PACS images. These three
images are also presented in Figure~\ref{fig:mosaic} along with the optical
image of NGC\,6888. 
These combinations of images exhibit in great detail the
nebular and dust distribution of NGC\,6888.  
The images confirm the suggestion of \cite{Toala2014} that NGC\,6888 is 
expanding towards a low-density region along the NW direction, which might 
be producing the expansion of the blowout. 
At the same time, the colour-composite IR pictures in Figure~\ref{fig:mosaic} 
indicate that NGC\,6888 is placed behind a molecular filament.

The colour-composite IR panels of Figure~\ref{fig:mosaic} disclose 
an interesting feature. In the three IR panels, there is a dark 
region separating the contribution from NGC\,6888 with that of 
the ISM. This effect is more clearly seen around the NE cap. 
The lack of IR emission is spatially coincident with the [O~{\sc iii}] 
optical emission, which according with \citet{Gruendl2000} traces the expanding shock generated by the wind-wind 
interaction that created the WR nebula. 
The diminished IR emission in this layer suggests that dust is being
destroyed by the expansion of the shock into the ISM. To further 
illustrate this, we present in Appendix~\ref{sec:appA} close up images of the
NE cap of NGC\,6888.

\section{IR photometry}

\begin{table*}
\centering
\setlength{\columnwidth}{0.1\columnwidth}
\setlength{\tabcolsep}{1.0\tabcolsep}
\caption{IR fluxes of NGC 6888 extracted from different regions.}
\begin{tabular}{ccccccccccc}
\hline
\multicolumn{1}{c}{Instrument} &
\multicolumn{1}{c}{$\lambda_\mathrm{c}$} &
\multicolumn{1}{c}{Complete} & 
\multicolumn{1}{c}{Region 1} & 
\multicolumn{1}{c}{Region 2} & 
\multicolumn{1}{c}{Region 3} & 
\multicolumn{1}{c}{Region 4} & 
\multicolumn{1}{c}{Region 5} &
\multicolumn{1}{c}{Region 6} & 
\multicolumn{1}{c}{NGC\,6888}\\
\multicolumn{1}{c}{} &
\multicolumn{1}{c}{($\mu$m)} &
\multicolumn{1}{c}{(Jy)} &
\multicolumn{1}{c}{(Jy)} & 
\multicolumn{1}{c}{(Jy)} & 
\multicolumn{1}{c}{(Jy)} & 
\multicolumn{1}{c}{(Jy)} & 
\multicolumn{1}{c}{(Jy)} & 
\multicolumn{1}{c}{(Jy)} &
\multicolumn{1}{c}{(Jy)} \\
\hline
{\it WISE}    & 12   & $17.3\pm6.3$    & $1.6\pm0.6$   &  $1.6\pm0.8$   & $1.7\pm1.1$   & $0.5\pm0.3$  & $1.9\pm0.5$   & $0.3\pm0.2$  & $9.8\pm2.7$    \\
{\it Spitzer} MIPS  & 24   & $66.9\pm11.4$   & $3.7\pm0.7$   & $4.1\pm1.0$    & $10.9\pm1.4$  & $1.7\pm0.3$  & $10.2\pm0.7$  & $0.2\pm0.3$  & $53.6\pm3.5$   \\
{\it Herschel} PACS  & 70   & $281.1\pm107.2$ & $32.9\pm8.2$  & $24.0\pm12.2$  & $51.0\pm16.3$ & $17.7\pm3.8$ & $57.7\pm7.8$  & $2.7\pm3.4$  & $302.5\pm41.0$ \\
{\it Herschel} PACS  & 100  & $320.5\pm187.9$ & $47.3\pm12.6$ & $59.1\pm18.5$  & $59.8\pm24.9$ & $24.3\pm5.8$ & $61.1\pm11.9$ & $13.4\pm5.0$ & $320.5\pm62.7$ \\
{\it Herschel} PACS  & 160  & $414.4\pm220.0$ & $47.5\pm13.2$ & $111.2\pm19.4$ & $73.3\pm26.1$ & $21.7\pm6.1$ & $26.8\pm12.5$ & $32.4\pm5.3$ & $140.8\pm65.6$ \\
\hline
\end{tabular}
\label{tab:fluxes}
\vspace{0.15cm}
\end{table*}

In order to produce a consistent model of the nebular and dust 
properties of NGC\,6888, we extracted the IR photometry 
of each available image presented in Figure~\ref{fig:panelIR} 
for $\lambda \geqslant 12$~$\mu$m. 
We did not use the {\it Spitzer} IRAC observations because they include the 
contribution from several nebular lines as well as possible low-ionization 
emission lines from the ISM. 
Moreover, the large number of 
background stars hampers an appropriate selection
of the extraction region.

The photometry extraction procedure is described in Paper~I. 
We define extraction regions where the integrated flux can be easily 
computed by adding all the pixels in the image.  
The contribution from background stars is excised and thus it is not taken
into account for the flux estimate. 
A similar procedure would be performed for background regions, but it shall be noted 
that the ISM around NGC\,6888 is not homogeneous as revealed by the IR images of 
NGC\,6888 shown in the previous section.    
To minimise the impact of the spatially-varying background, several background 
regions surrounding NGC\,6888 were selected to compute the same number of 
background-subtracted fluxes. 
We emphasize that regions SE from NGC\,6888 were discarded because they show the 
strongest contribution from the ISM according to Figures~\ref{fig:mosaic} and 
\ref{fig:panelIR}.

The resultant mean and standard deviation of the flux values 
are used as the flux and
its uncertainty ($\sigma_\mathrm{back}$), respectively. Nevertheless, 
a total error must be computed by adding 
other uncertainties such as that obtained from the 
callibration ($\sigma_\mathrm{cal}$) which depends
on each instrument. Thus, the total error can be 
calculated as
\begin{equation}
\sigma_\mathrm{tot} = \sqrt{\sigma_\mathrm{back}^2 + \sigma_\mathrm{cal}^2}
\end{equation}

\begin{figure}
\begin{center}
\includegraphics[width=\linewidth]{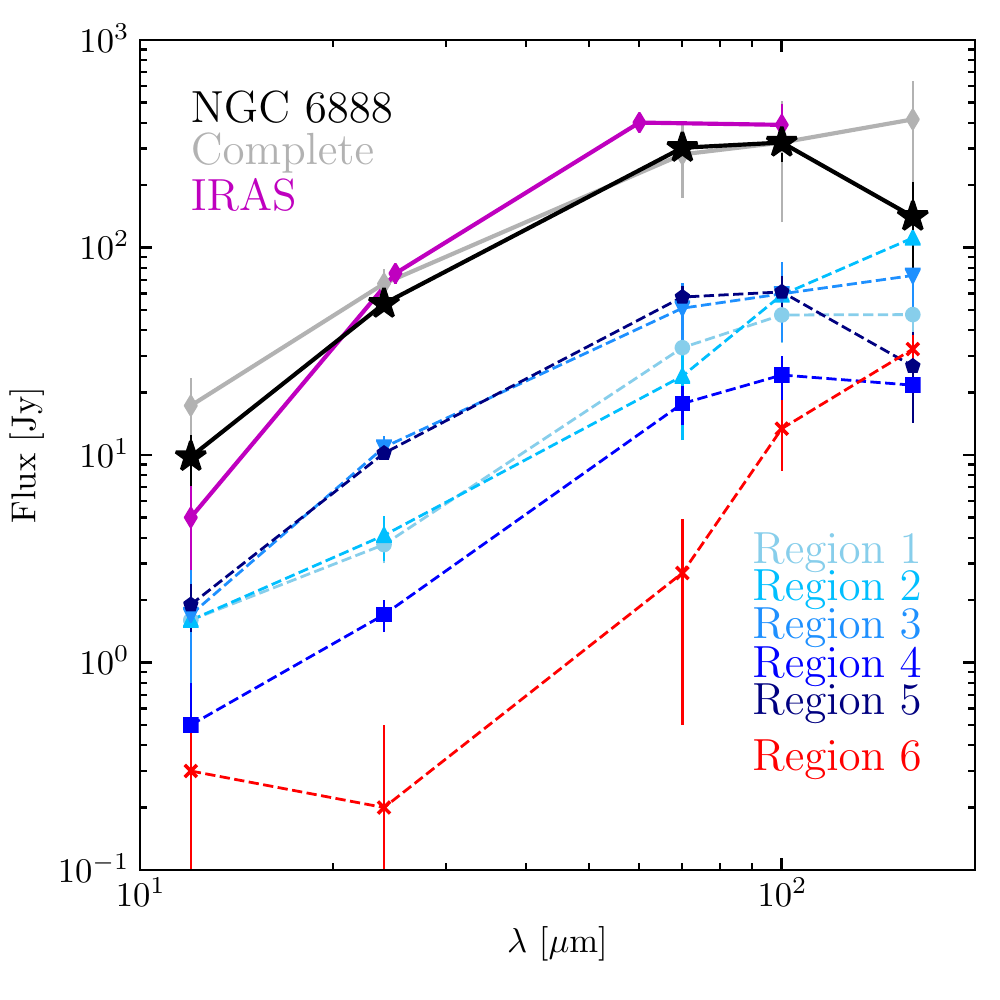}
\caption{SED of the WR nebula NGC\,6888 (stars) modeled in the present
paper. Different symbols (colours) 
show the SED of different regions in NGC\,6888 as defined in the 
bottom-central panel of Figure~\ref{fig:panelIR}. The SED labelled as 
Complete has been obtained taking into account the whole optical extension
of the nebula. The purple SED labelled as IRAS is the photometry used by 
\citet{Mathis1992}.}
\label{fig:sed}
\end{center}
\end{figure}

\begin{figure*}
\begin{center}
\includegraphics[width=\linewidth]{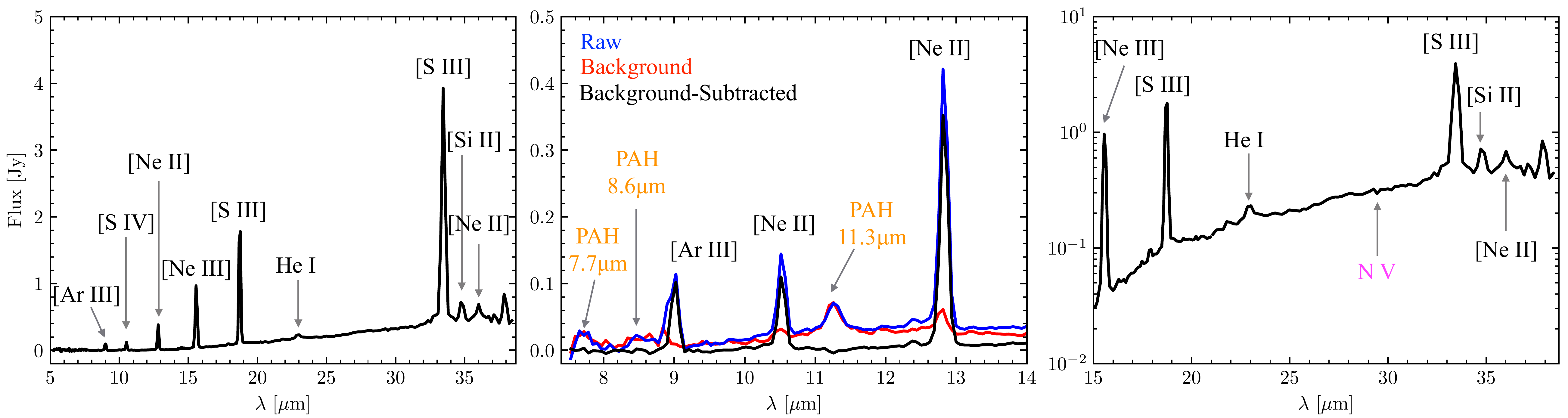}
\vspace{0.1in}
\caption{Background-subtracted low-resolution IRS spectra (black lines) extracted 
from Slit~4 as shown in Figure~\ref{fig:low_res_positions}. The most 
prominent lines are labelled. The left panel shows the complete 
range of the low-resolution spectra while the other two shown different regions
of the spectra. The middle panel also shows the background and background-unsubtracted
(raw) spectra.}
\label{fig:irs_spectra}
\end{center}
\end{figure*}

We first extracted the corresponding photometry from an aperture encompassing the 
emission from the [O~{\sc iii}] line which defines the complete nebular extension 
of NGC\,6888 (see Fig.~\ref{fig:mosaic}). 
This aperture is shown as a white dashed-line region in most panels of 
Figure~\ref{fig:panelIR}. 
The resultant flux and error values of this region are listed in 
Table~\ref{tab:fluxes} and illustrated in Figure~\ref{fig:sed} 
as a SED labelled as 
Complete. 
Figure~\ref{fig:sed} suggests that the Complete photometry of NGC\,6888 
peaks at wavelengths longer than 160~$\mu$m. 
However, as shown in Figure~\ref{fig:panelIR} the dashed-line region includes a 
significant contribution (depending on the IR image) from gas that very likely 
has a ISM origin. 

To assess the contribution from the ISM in the SED, we explored the variation 
of the IR SED from different regions within NGC\,6888. 
The bottom-central panel of Figure~\ref{fig:panelIR} shows the different 
extraction regions here considered.  
Their corresponding photometry is listed in Table~\ref{tab:fluxes} and 
plotted and compared to the Complete SED in Figure~\ref{fig:sed}. 
This figure shows that the IR SED extracted from Regions 1--5 exhibit 
significant differences, but these can be explained by considering the SED of 
Region~6.The SED of this region, defined from an external filament to NGC\,6888,
increases with wavelength, suggesting the presence of colder dust in the 
ISM (see Section~\ref{sec:dust_modelling}). Since the 
emission from NGC\,6888 is projected against that of the ISM, 
this effect should also be present in other regions with different relative 
contributions.  
For example, the SED of Region~2, defined at the SE edge of NGC\,6888, peaks 
at long wavelengths, i.e., it shows noticeable contribution from the ISM as 
shown by Region~6.

An inspection of the SED from these different regions led us to conclude that 
Region~5 has a SED (mostly) free from the contribution from the ISM. 
Its SED has a peak located around $\sim$70--100~$\mu$m consistent 
with the IR emission found in other WR nebulae. 
Assuming that the SED of Region~5 is thus representative of the shape of 
the SED of NGC\,6888, its flux level has been normalised to that of the 
emission from the whole nebula to create a SED labelled as NGC\,6888 
that would describe our target. 
The resultant SED, which we will use to model the 
dust properties of NGC\,6888, is presented in Figure~\ref{fig:sed} 
and the corresponding flux values are listed in the last 
column of Table~\ref{tab:fluxes}. For comparison we also show 
in Figure~\ref{fig:sed} the {\it IRAS} SED used 
in \citet{Mathis1992}. The latter has larger fluxes at mostly all
wavelengths except at 12~$\mu$m. 
The excess of the {\it IRAS} photometry compared to that estimated in this work 
is very likely due to the contribution from material in the line of sight and 
the point-like sources unresolved in the {\it IRAS} observations.

\begin{table*}
\setlength{\columnwidth}{0.2\columnwidth}
\setlength{\tabcolsep}{\tabcolsep}
\caption{Fluxes of the emission lines detected in the {\it Spitzer} IRS observations of NGC\,6888. 
Labels from 1 to 7 represent the extraction regions defined in Figure~\ref{fig:low_res_positions}.}

\begin{tabular}{lcccccccccl}

\hline
 Line & $\lambda$ &   &   &   & Flux (mJy) &   &   &   & Model \\
      & ($\mu$m)  & 1 & 2 & 3 & 4          & 5 & 6 & 7 &  (mJy) \\

\hline
[Ar~{\sc iii}] & 8.98 & 1.1$\pm$0.2  & 2.3$\pm$0.8   & 6.7$\pm$4.8 & 8.0$\pm$2.1 & 3.9$\pm$3.3 & 0.5$\pm$1.9 & 0.3$\pm$1.1 & 1.9 \\
$[$S~{\sc iv}$]$ & 10.5 & 0.2$\pm$0.05 & 0.7$\pm$0.3 & 5.6$\pm$3.9 & 8.4$\pm$2.2 & 3.1$\pm$2.7 & 0.4$\pm$1.6 & 0.1$\pm$0.5 & 2.2 \\
$[$Ne~{\sc ii}$]$ & 12.8 & 5.2$\pm$0.1 & 4.3$\pm$0.5 & 5.4$\pm$1.4 & 6.1$\pm$0.7 & 3.3$\pm$1.4 & 0.9$\pm$0.9 & 1.3$\pm$1.0 & 1.4  \\
$[$Ne~{\sc iii}$]$ & 15.5 & 2.9$\pm$0.004 & 3.1$\pm$0.04 & 6.4$\pm$0.05 & 10.8$\pm$0.03 & 3.0$\pm$0.005 & 1.2$\pm$0.03 & 1.0$\pm$0.05 & 2.5 \\
$[$S~{\sc iii}$]$ & 18.7 & 8.6$\pm$0.01 & 9.3$\pm$0.2 & 12.4$\pm$0.2 & 17.4$\pm$0.08 & 5.7$\pm$0.01 & 2.1$\pm$0.09 & 2.8$\pm$0.3 & 9.3 \\
$[$S~{\sc iii}$]$ & 33.5 & 19.1$\pm$0.02 & 20.2$\pm$0.04 & 26.1$\pm$0.04 & 34.9$\pm$0.05 & 13.0$\pm$0.03 & 5.8$\pm$0.03 & 6.7$\pm$0.05 & 22.6  \\
$[$Si~{\sc ii}$]$ & 34.8 & 2.9$\pm$0.02 & 2.4$\pm$0.02 & 2.8$\pm$0.02 & 2.9$\pm$0.03 & 1.8$\pm$0.02 & 
1.4$\pm$0.02 & 1.8$\pm$0.02 & 7.8 \\
\hline
$n_\mathrm{e}$([S\,{\sc iii}])  & [cm$^{-3}$] & 420 & 440 & 470 & 520 & 400 & 230 & 350 & 340 \\
\hline
\label{tab:irs}
\end{tabular}
\end{table*}

\section{IR spectra}

{\it Spitzer} IRS spectra of NGC\,6888 were extracted from the 
brightest  regions following 
the N-NW-SW direction. The location of the extraction regions, as shown in Figure~\ref{fig:low_res_positions}, are labelled as 
Slit\,1 to 7 and correspond to  regions with 
the lowest or negligible contribution from the ISM. The spectra 
are very similar with a continuum from dust emission. 
The dominant emission lines in these spectra are those of 
[Ne\,{\sc ii}] 12.8~$\mu$m, 
[Ne\,{\sc iii}] 15.5~$\mu$m, 
[S\,{\sc iii}] 18.7 and 33.5~$\mu$m, 
[S\,{\sc iv}] 10.5~$\mu$m, 
[Ar\,{\sc iii}] 8.98~$\mu$m, and
[Si\,{\sc ii}] 34.8~$\mu$m (see Fig.~5). 
Other less intense lines correspond to He\,{\sc i} at 22.9~$\mu$m and very likely an absorption due 
to N\,{\sc v} at 29.5~$\mu$m. 
The latter implies the presence of gas with temperature around 
$\sim$10$^{5}$~K at the mixing layer between the nebular cold
($\sim$10$^{4}$~K) and the hot X-ray-emitting gas ($\sim$10$^{6}$~K)
filling NGC\,6888 \citep[e.g.,][]{Gruendl2004,Fang2016}.
As an example, we show in Figure~\ref{fig:irs_spectra} the
spectrum from Slit~4 with the most prominent lines labelled. 

The IRS spectra can be used to look for the presence of 
molecules in NGC\,6888 such as H$_2$ and polycyclic aromatic hydrocarbons 
(PAHs) similarly to other nebulae around evolved stars such as planetary 
nebulae \citep[see, e.g.,][and references therein]{Mata2016,Fang2018,Toala2019} 
and, according to \citet{StLouis1998}, the interstellar nebula around WR\,7 
(NGC\,2359).
PAH features appear in the spectra of NGC\,6888, but, after subtracting the 
background spectra, these disappear as illustrated in the middle panel of 
Figure~\ref{fig:irs_spectra} showing the spectral range including the 7.7, 
8.6 and 11.3~$\mu$m PAH features. 
Furthermore, no H$_2$ lines are detected in the IRS spectra of NGC\,6888.

We used the {\sc pahfit} routines \citep{smith2007b} to measure the line 
intensities and errors for each line for all 7 spectra. The ions, 
flux and errors of the most prominent lines are listed in 
Table~\ref{tab:irs}. 
This table shows
clear variations in the measured lines from slit to slit. In particular,
emission lines from the spectrum 4 exhibit the largest
fluxes. We attribute these differences to the variations of the physical
properties within the slit positions, e.g., due to the clumpy distribution 
of gas in NGC\,6888. To corroborate this, we have computed
the electron densities ($n_\mathrm{e}$)
using the [S\,{\sc iii}]~18.7 and 33.5~$\mu$m 
with {\sc pyneb} \citep{Luridiana2015}. The $n_\mathrm{e}$([S\,{\sc iii}]) 
values are listed in the bottom row of Table~\ref{tab:irs} for each slit 
position and, indeed, show that Slit~4 has the largest density estimate.

\section{DUST MODELING}
\label{sec:dust_modelling}

\begin{figure}
\begin{center}
  \includegraphics[width=\linewidth]{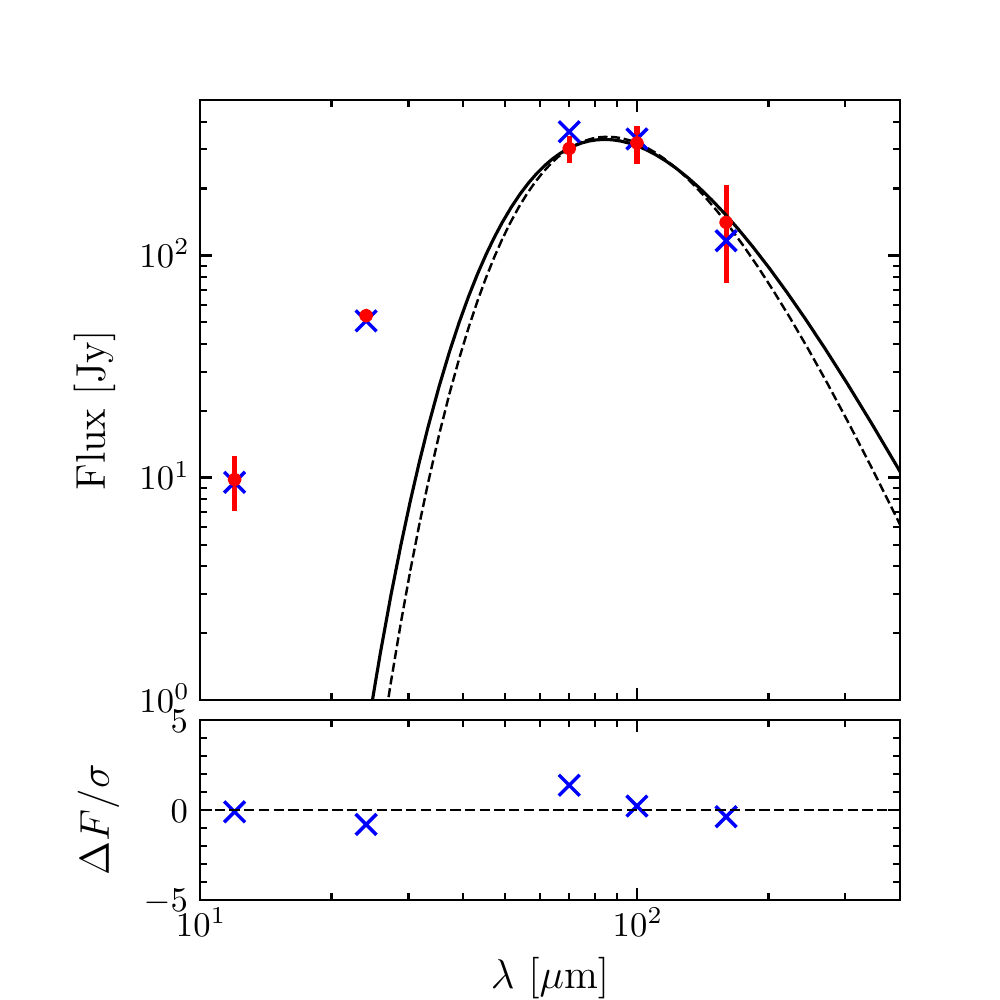}
\caption{Observed IR SED of NGC\,6888 (red dots) compared to the synthetic 
IR SED of our best fit model (blue crosses). The solid and dashed lines correspond
to a MBB model with fixed $\beta=2$ ($T_\mathrm{dust}$=34.4$\pm$0.8~K) and a MBB model with
$\beta$ as a free parameter 
($\beta$=2.7$\pm$0.6, $T_{\text{dust}}$=29.3$\pm$0.3~K), respectively. 
The bottom panel shows the
residuals between the observed and synthetic IR SED.}
\label{fig:sed_fit}
\end{center}
\end{figure}

As a first approximation to the dust temperature ($T_\mathrm{dust}$) and mass 
($M_\mathrm{dust}$), we fitted a modified blackbody (MBB) model to the IR SED of 
NGC\,6888 with $\lambda \geqslant 70$~$\mu$m. Under the assumption that dust 
is optically thin and that all the dust has a single temperature, the MBB
can be expressed as
\begin{equation}
F_\nu=\kappa_{\nu_0}\left(\frac{\nu}{\nu_0}\right)^{\beta}M_{\text{dust}}\frac{B_\nu(T_{\text{dust}})}{d^2},
\end{equation}
\noindent
where $B_\nu(T_{\text{dust}})$ is the Planck function and $\kappa_{\nu_0}$ is the dust emissivity adopted 
to be $\kappa_{\nu_0}=27.1\ \text{g}\ \text{cm}^{-2}$ at the 
reference wavelength $\lambda$=100~$\mu$m
\citep[][]{Draine2003}. We fitted the {\it Herschel} photometry taking $\beta$, $T_{\text{dust}}$ and $M_{\text{dust}}$ as free parameters. 
The best fit that minimises $\chi^{2}$ corresponds
to $\beta=2.7\pm0.6$, $T_{\text{dust}}=29.3\pm0.3\ \text{K}$ and $M_{\text{dust}}=0.69\pm0.04\ \text{M}_{\odot}$. Similar temperatures for WR nebulae have been previously estimated with {\it IRAS} observations \citep[e.g.,][]{Marston1991}.
However, there is a known degeneracy between $\beta$
and $T_\mathrm{dust}$ \citep{Juvela2013} and thus, we computed another
model by fixing $\beta=2$ that resulted in $T_{\text{dust}}=34.4\pm0.8\ \text{K}$ and 
$M_{\text{dust}}=0.32\pm0.04\ \text{M}_{\odot}$.  We note that
the MBB fits are not meant to reproduce the complete IR photometry of NGC\,6888, but 
they can be useful to estimate the temperature
of the large grain population, another constrain to our detailed model (see below). Nevertheless, these approximations 
reveal that the dust grains must be present in a wide range of sizes and temperatures.
These two
MBB models are plotted in Figure~\ref{fig:sed_fit} in comparison with 
the observed IR SED. 
To further illustrate the differences with the physical properties of the 
material in the surrounding ISM, we also estimated the dust temperature for 
the photometry extracted for Region~6 (see Figures~\ref{fig:panelIR} and 
\ref{fig:sed}) probing this is ISM. 
The dust temperature for this emission is lower, $T_\mathrm{ISM}$=16.7$\pm$0.4~K,  
as expected for cold ISM dust \citep[see][and references therein]{ostrovskii2020}.

Our goal is to produce a characterization of the dust properties in 
NGC\,6888 consistent with the nebular optical properties. For this, we used
the spectral synthesis and plasma simulation code {\sc Cloudy} 
\citep[version 17.01;][]{Ferland2017} coupled 
with the {\sc pyCloudy} libraries \citep{Morisset2013}.
The input parameters required by {\sc Cloudy} are i) the form of the 
incident spectrum (a model of the spectrum of WR\,136), 
ii) the density distribution, iii) abundances, and 
iv) the dust properties (chemical composition and size distribution). \\
With these, {\sc Cloudy} is able to 
produce an emission model which is subsequently processed with 
the help of {\sc pyCloudy} to create synthetic optical long-slit 
observations. Furthermore, synthetic IR SED photometry is produced by
using the transmission curves from the different IR instruments used 
here\footnote{The transmission curves of the {\it WISE}, {\it Spitzer} 
and {\it Herschel} instruments were obtained from \url{http://svo2.cab.inta-csic.es/theory/fps/index.php?mode=browse}}. 
The synthetic observations are then compared to previously published 
optical studies of NGC\,6888 and the IR observations (spectra and IR SED) 
presented in the previous sections to 
assess the validity of our models.

\begin{table}
\centering
\setlength{\columnwidth}{0.1\columnwidth}
\setlength{\tabcolsep}{1.0\tabcolsep}
\caption{Elemental abundances of NGC\,6888 in units of 
12$+$log(X/H) reported for the slit A6 in  
\citet{Esteban2016} (E2016), \citet{mesa2014} (MD2014) 
and those used by \citet{reyes2015} (RP2015). 
The last column shows the 
abundances adopted for our model.}
\label{tab:elemental_abundances}
\begin{tabular}{ccccc}
\hline
Element & E2016 & MD2014 & RP2015 & Model \\
\hline
He & 11.23$\pm$0.06 & 11.21$\pm$0.03 & 11.21 & 11.23  \\
O & 8.19$\pm$0.13 & 8.20$\pm$0.09 & 8.20 & 8.05 \\
C & -- & 8.86$\pm$0.31 & 8.86 & 8.86 \\
N & 8.27$\pm$0.18 & 8.54$\pm$0.20 & 8.40 & 8.68 \\
Ne & 7.51$\pm$0.78 & 7.51$\pm$0.20 & 7.51 & 7.78 \\
S & 6.92$\pm$0.21 & 6.77$\pm$0.20 & 7.10 & 6.92 \\
Cl & 4.99$\pm$0.15 & -- & -- & 4.84 \\
Ar & 6.41$\pm$0.11 & 6.41$\pm$0.11 & 6.41 & 6.41 \\
\hline
\end{tabular}
\end{table}

\begin{table}
\begin{center}
\setlength{\columnwidth}{0.1\columnwidth}
\setlength{\tabcolsep}{\tabcolsep}
\caption{Emission lines for region A6 of NGC\,6888 in \citet{Esteban2016}
compared to the predictions from our model. All line fluxes are
normalized with respect to H$\beta$=100.}
\begin{tabular}{cccc}
\hline
\multicolumn{1}{c}{Line} &
\multicolumn{1}{c}{$\lambda$} &
\multicolumn{1}{c}{\citet{Esteban2016}} &
\multicolumn{1}{c}{Model}\\
\multicolumn{1}{c}{} &
\multicolumn{1}{c}{$($\AA$)$} &
\multicolumn{1}{c}{(Slit A6)} &
\multicolumn{1}{c}{}\\
\hline
$[$O~{\sc ii}$]$    & 3726   & 50.4$\pm$4.7 & 39.4 \\
$[$O~{\sc iii}$]$  & 4363   & 1.4$\pm$0.4 & 1.5 \\
He I   & 4471   & 8.8$\pm$0.4 & 8.1 \\
$[$O~{\sc iii}$]$   & 4959   & 81.3$\pm$1.7 & 82.1 \\
$[$O~{\sc iii}$]$ & 5007 & 242.9$\pm$4.9 & 245.1 \\
$[$Cl~{\sc iii}$]$ & 5518 & 0.5$\pm$0.1 & 0.6 \\
$[$Cl~{\sc iii}$]$ & 5538 & 0.4$\pm$0.1 & 0.4 \\
$[$N~{\sc ii}$]$ & 5755 & 1.3$\pm$0.1 & 2.2 \\
He I & 5876 & 25.8$\pm$0.7 & 23.6 \\
$[$N~{\sc ii}$]$ & 6548 & 41.8$\pm$1.2 & 51.0 \\
H$\alpha$ & 6563 & 301.6$\pm$9.0 & 286.8 \\
$[$N~{\sc ii}$]$ & 6583 & 130.3$\pm$3.9 & 150.3 \\
He I & 6678 & 7.0$\pm$0.4 & 5.9 \\
$[$S~{\sc ii}$]$ & 6716 & 5.1$\pm$0.3 & 4.4 \\
$[$S~{\sc ii}$]$ & 6731 & 4.2$\pm$0.3 & 3.6 \\
He I & 7065 & 4.2$\pm$0.5 & 4.2\\
$[$Ar~{\sc iii}$]$ & 7136 & 15.2$\pm$0.7 & 26.1 \\
\hline
$c$(H$\beta$) &  & 0.62$\pm$0.03 & 0.65 \\
log($F$(H$\beta$)) & [erg s$^{-1}$ cm$^{-2}$]  & $-$13.01$\pm$0.01 & $-$13.01 \\

$n_\mathrm{e}$ ($[$S~{\sc ii}$]$) & [cm$^{-3}$]   &  200$\pm$140  &  198 \\
$T_\mathrm{e}$ ($[$O~{\sc iii}$]$) & [K]   & 9550$\pm$820 & 9708 \\
\hline
\end{tabular}
\label{tab:emmision_lines}
\vspace{0.15cm}
\end{center}
\end{table} 

We started our modeling adopting the most recent abundances determinations
reported by \citet[][]{Esteban2016}, which correspond to 10~m Gran Telescopio 
de Canarias (GTC) OSIRIS spectroscopic observations. 
Their slit A6 is placed on a region spatially coincident with the one used here to 
estimate the total IR photometry of NGC\,6888 
\citep[see Figure~1 in][]{Esteban2016}. 
Their abundances and emission line fluxes obtained from these observations are 
listed in Tables~\ref{tab:elemental_abundances} and \ref{tab:emmision_lines}, 
respectively. The total 
H$\beta$ flux ($F(\mathrm{H}\beta)$) in logarithmic scale, the electron 
temperature ($T_\mathrm{e}$) and $n_\mathrm{e}$ obtained from the 
[O\,{\sc iii}] and [S\,{\sc ii}] lines, respectively, are also presented in the
bottom rows of Table~\ref{tab:emmision_lines}. 
We note that during the fitting process, some of the
abundances had to be tuned in order to reproduce simultaneously the optical and 
IR emission lines. For comparison we also show in Table~\ref{tab:elemental_abundances}
the abundances reported by other works.

\begin{table}\centering
\setlength{\columnwidth}{0.1\columnwidth}
\setlength{\tabcolsep}{1.0\tabcolsep}
\caption{Principal stellar parameters of WR 136. \label{tab:stellar_parameters}}
\begin{tabular}{lcl}
\hline
$d$ [kpc]                & 1.9$^{+0.14}_{-0.12}$    & \citet{BJ2018}    \\
WR-subtype               & WN6h   & \citet{Hamann2019}\\
$M_\star$ [M$_{\odot}$]  & 23/21  & \citet{Hamann2019}\\
log($L$) [L$_{\odot}$]   & 5.78   &\citet{Hamann2019}\\
$T_{\star}$ [kK]         & 70.8   & \citet{Hamann2019}\\
\hline
\end{tabular}
\end{table}

\begin{figure}
\begin{center}
  \includegraphics[width=\linewidth]{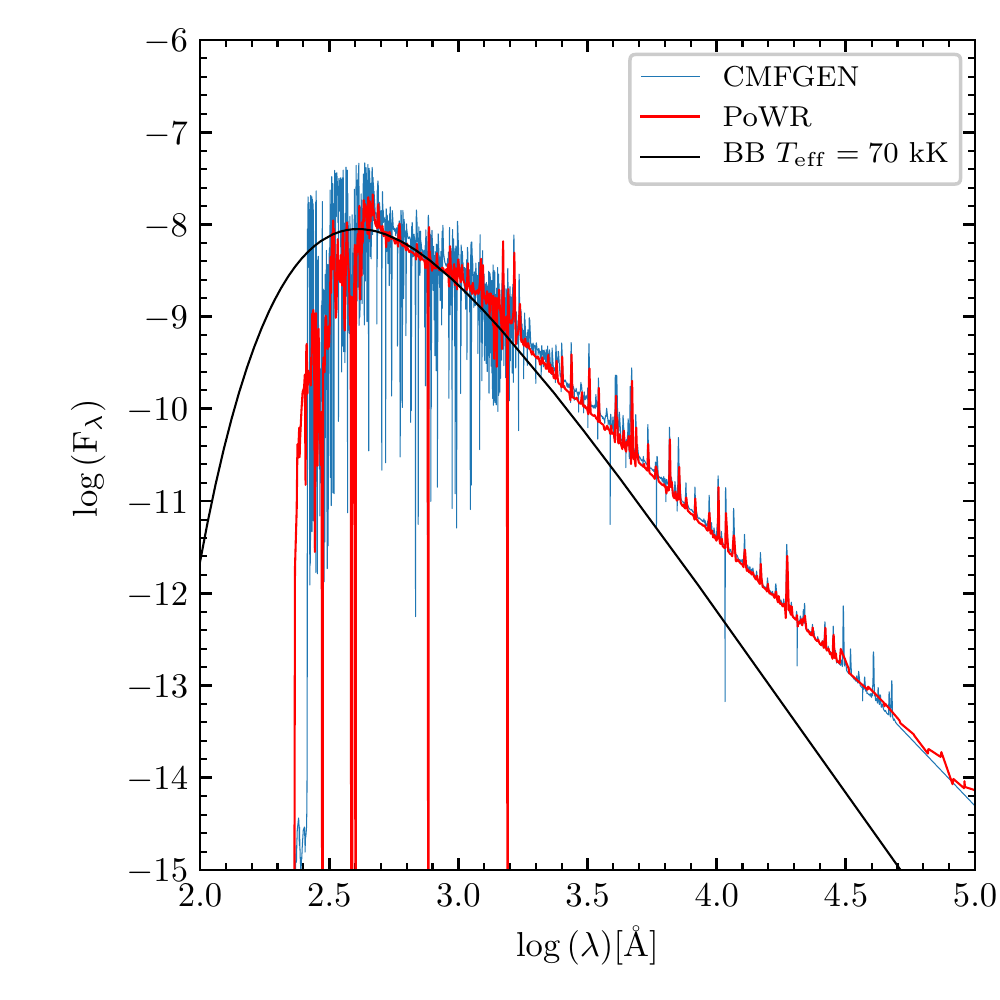}
\caption{Comparision between different stellar atmosphere models. 
The blue line correspond to the stellar atmosphere from {\sc cmfgen} used
in \citet{reyes2015} while the red line is the PoWR model used in the
present work. The black line presents a blackbody emission model with 
an effective temperature of 70~kK.}
\label{fig:sed_comparation}
\end{center}
\end{figure}

We demonstrated in Paper~I that 
a detailed prescription of the stellar atmosphere
is necessary to obtain a more 
realistic description of the nebular
and dust parameters. The progenitor star of NGC\,6888, 
WR\,136, has been modeled by \citet{Hamann2019} using the updated version of the 
stellar atmosphere code PoWR. 
The best fit stellar parameters reported in that
work are listed in Table~\ref{tab:stellar_parameters}. Accordingly, we retrieved the model
labelled as WNL 10-16 with Galactic metallicity and hydrogen fraction of 0.2 
from the PoWR database. This model is presented in 
Figure~\ref{fig:sed_comparation} compared to
a blackbody emission model with the same effective temperature (70 kK) as well as the stellar model used by \citet{reyes2015} kindly provided by J.\,Reyes-P\'{e}rez. In the following, 
we will adopt the distance of $d=1.9^{+0.14}_{-0.12}$~kpc 
estimated by \citet{BJ2018} using the {\it Gaia} data release.

\subsection{Detailed modeling of NGC\,6888}

The H$\alpha$ and [N\,{\sc ii}] images of NGC\,6888 show an apparent elliptical 
shape with semi-major and semi-minor axes of 540$^{\prime\prime}$ and 
360$^{\prime\prime}$, respectively.  
At a distance of 1.9~kpc, the physical size of NGC\,6888 is 5~pc and 3.9~pc, respectively. 
\citet{Marston1988} showed however that the kinematics of NGC\,6888 can be
broadly fit by an expanding spherical shell with an expansion velocity of 
85~km~s$^{-1}$, although we note that a shell with blisters would be more 
accurate as unveiled by the [O\,{\sc iii}] image.
For this, we adopted a spherically-symmetric shell for our model. 
An averaged angular 
radius of 450$^{\prime\prime}$ was set as a fixed parameter for the outer 
radius ($r_\mathrm{out}$). The inner radius ($r_\mathrm{in}$) was varied 
along with the filling factor ($\epsilon$) and $n_\mathrm{e}$ to try to
fit the optical observations and physical properties of NGC\,6888.
Values between $n_\mathrm{e} \lesssim 100-500$~cm$^{-3}$ 
for the electron density have been reported in the literature 
\citep[][see also Table~\ref{tab:irs}]{Esteban1992,fernandez2012,Esteban2016}, 
but we note that it is very likely 
that the largest values correspond to dense clumps detected
in H$\alpha$ and [N\,{\sc ii}] narrow-band images (see Fig.~1).

\begin{figure}
\begin{center}
  \includegraphics[width=0.9\linewidth]{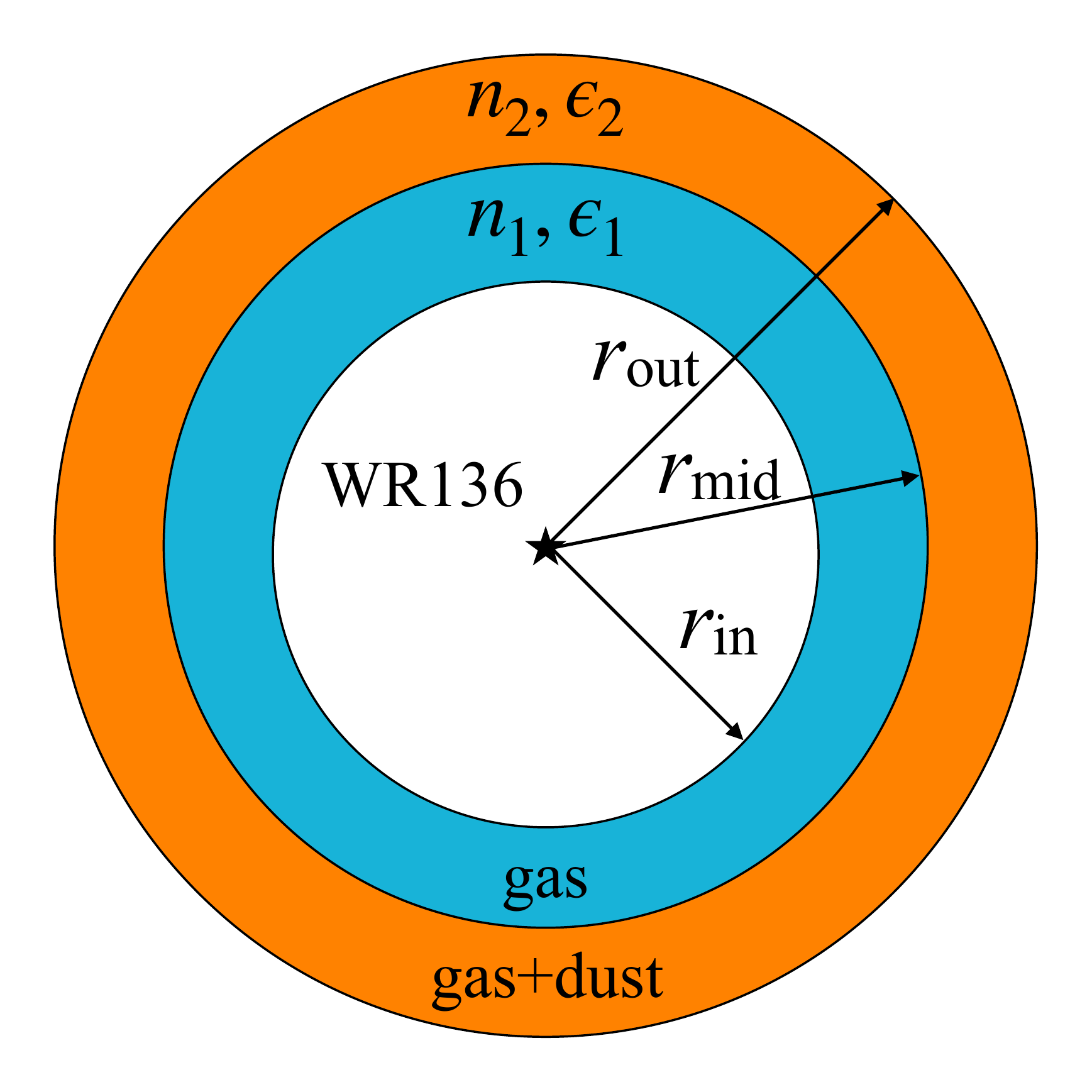}
\caption{Schematic view of the two-shell density distribution 
used to model the nebular and IR
properties of NGC\,6888. The outer shell 
accounts for the presence of gas
and dust while the inner shell only contains gas.}
\label{fig:esquema}
\end{center}
\end{figure}

We performed a large number of models
fixing $r_\mathrm{out}$ but varying $r_\mathrm{in}$ for families of $n_\mathrm{e}$ values. None of these models resulted in good nebular parameters comparable
to those reported for NGC\,6888. 
Consequently, we followed the results from \citet{reyes2015} and considered 
two shells: an inner shell with density $n_1$ between $r_\mathrm{in}$ and $r_\mathrm{mid}$ and a filling factor $\epsilon_1$
composed purely of gas, and an outer shell with density $n_2$ between 
$r_\mathrm{mid}$ and $r_\mathrm{out}$ and $\epsilon_2$. 
The outer shell includes contributions from both gas and dust. 
A schematic view of this model is presented in Figure~\ref{fig:esquema}.

Our best model that reproduces the nebular properties 
of NGC\,6888 was achieved with $n_1=400$~cm$^{-3}$, 
$n_2=180$~cm$^{-3}$, $\epsilon_1=10^{-3}$, $\epsilon_2=0.07$. The radii are
$r_\mathrm{in}=$400$^{\prime\prime}$, $r_\mathrm{mid}=$425$^{\prime\prime}$, 
and $r_\mathrm{out}=$450$^{\prime\prime}$. 
The synthetic intensities of 
the most important optical lines are listed in 
Table~\ref{tab:emmision_lines} in
comparison with those reported in \citet{Esteban2016} 
for their slit A6. 
The emission IR lines obtained from our model are compared to those obtained 
from the {\it Spitzer} IRS observations in Table~\ref{tab:irs}.
Despite the variations obtained from the {\it Spitzer} IRS spectra, the model 
broadly reproduces the IR lines, corroborating previous findings that suggest 
that the inner regions of NGC\,6888 are dominated by photoionization \citep{Esteban1992,Esteban1993,reyes2015}. 
Although we note that there might be certain 
components associated to shocks which produce line excitation \citep{moore2000,Gruendl2000}.

As noted before, we started our fitting process by adopting the set of 
abundances reported for slit A6 by \citet{Esteban2016}.  
However, some elemental abundances required small changes in order to fit 
both the optical and IR emission lines.
Nevertheless, most of them resulted in values close to those reported in 
the literature. 
In particular, the O abundance had to be reduced to 8.05, a small value 
compared to what has been reported by the references in 
Table~\ref{tab:elemental_abundances}, but marginally similar to the lower 
value reported by \citet{Esteban2016}.
The last column in Table~\ref{tab:elemental_abundances} lists the final 
abundances used for our best model.

\begin{figure}
\begin{center}
  \includegraphics[width=\linewidth]{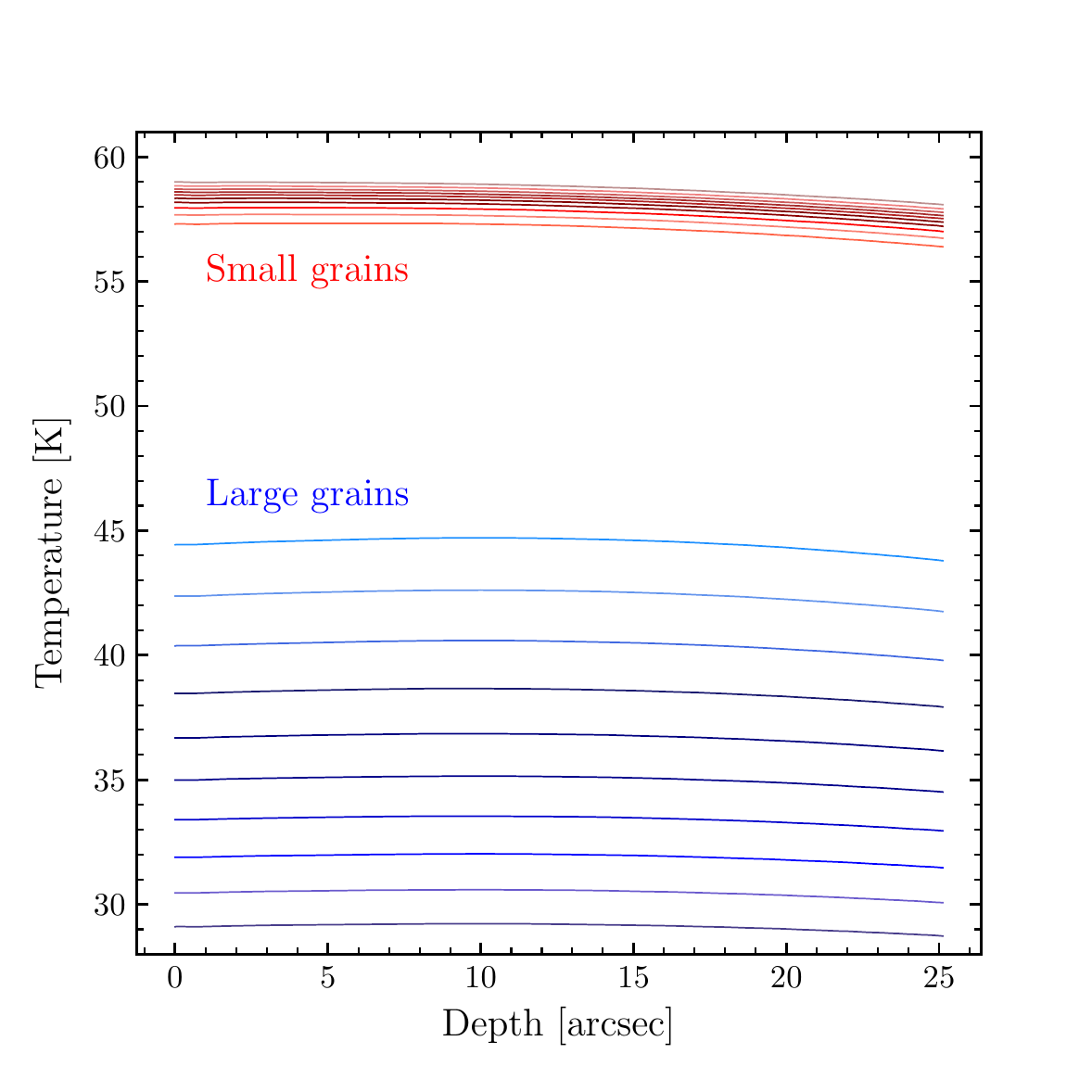}
\caption{Temperature distribution for the different dust sizes used for our best model of NGC\,6888. Each line represents a size bin of the small ($a_\mathrm{small}$=[0.002--0.008]~$\mu$m) and large grains 
($a_\mathrm{big}$=[0.05--0.5]~$\mu$m).}
\label{fig:grain_temperature}
\end{center}
\end{figure}

Once we have modeled the observed emission lines of NGC\,6888 we now include 
dust in our calculations to produce synthetic IR photometry. 
During the RSG and LBV phase, massive stars are copious producers of oxygen 
which condensate molecules, such as silicates in their atmospheres \citep{Dhar2020,Ver2009}. 
Accordingly, we adopt grains composed  of amorphous astronomical silicate. 
In particular, Cloudy includes olivine (MgFeSiO$_4$) as the default silicate grain.

Initially, we used two distribution of dust sizes as those used in 
\citet{Mathis1992}: a distribution for small
grains with sizes $a_\mathrm{small}$=[0.002--0.008]~$\mu$m and another 
for big grains with sizes $a_\mathrm{big}$=[0.005--0.25]~$\mu$m. 
However,
our first attempts to model the IR SED suggested the need of extending the 
size of the large grains. Finally, we found that a population of sizes for the
big grains of $a_\mathrm{big}$=[0.05--0.50]~$\mu$m improved our model.
The maximum size of the big grains was fixed to 0.50~$\mu$m as found for the
RSG star VY Canis Majoris \citep{Scicluna2015}. In all cases, the grains are spherical and their sizes follow a \citet[][]{Mathis1977} power-law distribution $\propto a^{-3.5}$ with 10 size bins for each distribution.

To reproduce the observed IR SED, our model requires a large amount of big 
grains, with a big grains-to-small grains ratio of 16:1 and a resultant 
dust-to-gas ratio of $5.6\times 10^{-3}$. 
The synthetic photometry obtained from our model is compared
to the observed IR SED in Figure~\ref{fig:sed_fit}. 
Our best model suggests that the total mass of NGC\,6888 is 
$25.5^{+4.7}_{-2.8}$~$M_{\odot}$ with a total dust mass of 
$M_\mathrm{dust}=0.14^{+0.03}_{-0.01}\ M_{\odot}$, less than a half of that estimated by the
MBB model with fixed $\beta=2$. The model predicts a mass for the big grains of
0.134$^{+0.03}_{+0.01}$~M$_{\odot}$. 
We note that the errors in the total mass and dust mass estimates 
were obtained by propagating the errors in the distance as reported by \citet{BJ2018}.

In Figure~\ref{fig:grain_temperature} we show the temperature 
distribution for different bin sizes of the dust grains 
located in the outer shell. 
The temperature for the large
grains is in the range estimated using the MBB approximation.

\section{Discussion} 

\subsection{On the spatial distribution of dust in NGC\,6888}

Our analysis above has disclosed an unprecedented view of the IR emission of the WR nebula 
NGC\,6888. The {\it Herschel} PACS and {\it Spitzer} IRAC images have revealed 
in great detail the distribution of dust in this WR nebula and the ISM around it. 
These are shown in the colour-composite IR panels of Figure~\ref{fig:mosaic}
which are compared to the optical image of NGC\,6888. Such detailed 
morphological IR characteristics of NGC\,6888 are only allowed by the 
spatial resolution of the observations used here as compared 
with the {\it IRAS} images used in previous analyses \citep[][]{Marston1991,Mathis1992}.

An interesting feature seen in the IR panels in Figure~\ref{fig:mosaic}
is a gap between the dust in NGC\,6888 and the emission that 
traces the outer cold dust from the ISM. This situation is clearly seen 
in the NE cap of NGC\,6888
and to a lesser extent towards the SW cap. 
For example, Figure~\ref{fig:mosaic} bottom right panel 
shows a dark region separating the {\it Herschel} images at 70~$\mu$m and 
160~$\mu$m. 
A careful comparison between the [O\,{\sc iii}] and IR images of the nebula 
shows that the former is distributed exactly in the dark region observed in 
the colour-composite IR panels of this figure (see Appendix~\ref{sec:appA}).

\begin{figure*}
\begin{center}
\includegraphics[width=0.44\linewidth]{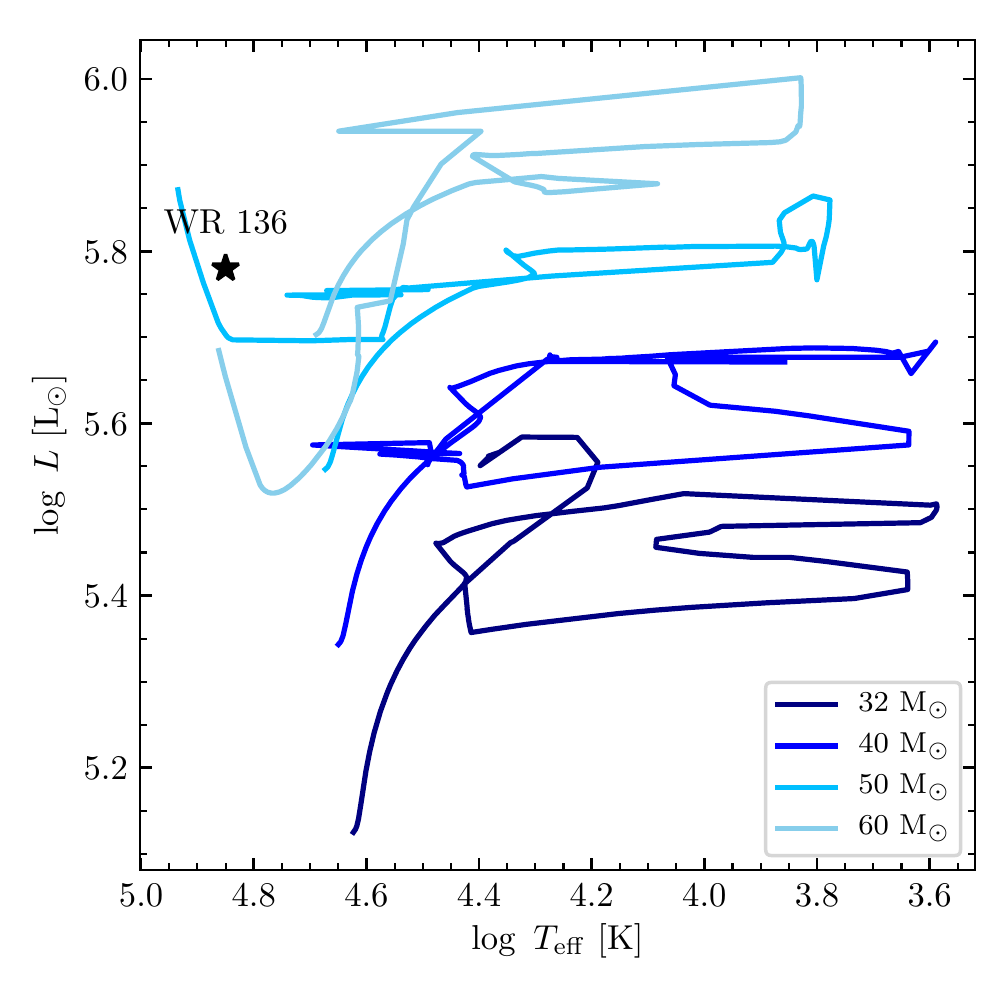}~
\includegraphics[width=0.44\linewidth]{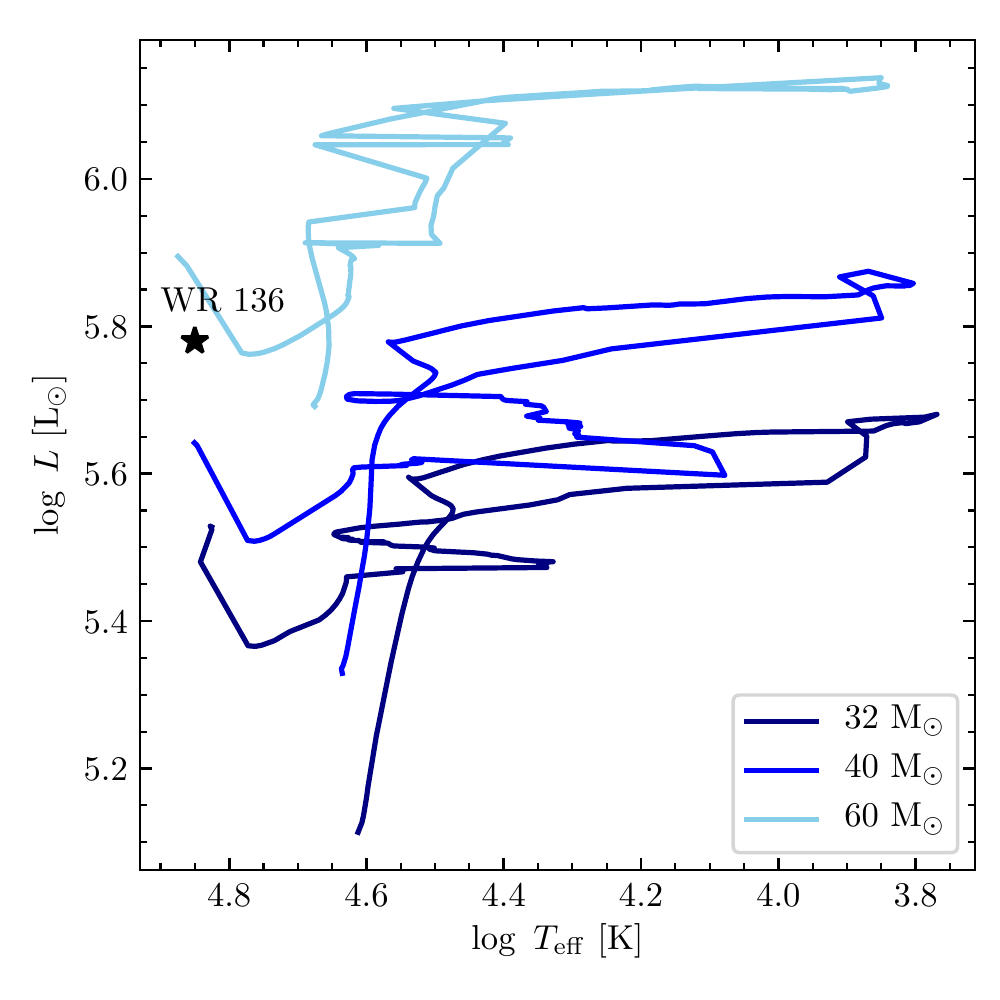}
\caption{Hertzsprung–Russell diagrams of stellar evolution models with initial
masses between 30 and 60 M$_\odot$ from \citet{Ek2012}. 
The black star represents the position of WR\,136. Different
colours represent models from different initial masses. The left (right) panel
presents model without (with) stellar rotation. There is no star model available
with initial mass of 50 M$_\odot$ in the Geneva Code web page.}
\label{fig:tracks}
\end{center}
\end{figure*}

\citet{Gruendl2000} presented the analysis of H$\alpha$ and [O\,{\sc iii}] 
narrow-band images of a sample of 8 bright WR nebulae, including NGC\,6888. 
They showed that 
the [O\,{\sc iii}] emission is smooth and traces the leading shock of the 
expanding nebula,
while the H$\alpha$ emission displays a clumpy morphology trailing inside
the [O\,{\sc iii}] emission (see Fig.~\ref{fig:mosaic} top left panel). 
\citet{Gruendl2000} argued that the displacement 
between these two emission lines is due to radiative cooling. 
Behind the shock front, the temperature drops while the density 
increases displacing the two emissions. 
We suggest that this dark region, which 
is coincident with the [O\,{\sc iii}] 
leading shock, is unveiling the destruction of dust 
at the edge of NGC\,6888.

The destruction of dust grains may be due to thermal evaporation, 
gas-grain collisions and destruction by grain-grain collisions 
\citep[see][]{woitke1993}. 
According to \cite{jones1996} the destruction of silicate dust 
grains in low velocity shocks can be attributed to non-thermal sputtering while
large grains ($\sim$0.1~$\mu$m) are easily shattered in 
grain-grain collisions. \citet{Slavin2004} presented calculations on the dust-gas 
decoupling on dust destruction in shocks. They found that grains with a large
range of sizes are almost completely destroyed by shock waves with velocities
around 75--150~km~s$^{-1}$. The reported expansion velocity of NGC\,6888, 
$\sim$80~km~s$^{-1}$ \citep[][]{Chu1983,Marston1988}, makes 
plausible the destruction of dust at the edge of NGC\,6888 
as the nebula expands. 

\subsection{Consequences of the dust model}

Taking advantage of an improved analysis of the spatial distribution of dust in
NGC\,6888, we managed to uncover its true IR SED. This represents an improvement
over the nebular estimates made from low-resolution {\it IRAS} observations. 
Our best model is able to reproduce nebular optical and IR properties of NGC\,6888 using 
an appropriate description of the stellar atmosphere of WR\,136. This resulted 
in a total mass of NGC\,6888 of 25.5$^{+4.7}_{-2.8}$~M$_{\odot}$ with a dust-to-gas 
ratio of $5.6\times10^{-3}$,
that is, a total dust mass of $M_\mathrm{dust}=$0.14$^{+0.03}_{-0.01}$~M$_{\odot}$. 
If we assume a duration of $2\times10^{5}$~yr for the RSG phase, we can estimate 
an averaged mass-loss rate of $(1.28^{+0.23}_{-0.15})\times10^{-4}$~M$_{\odot}$~yr$^{-1}$, consistent with that 
expected for the RSG phase.

Our results represent an improvement to the work of \citet{Mathis1992}.
They tried to fit
simultaneously the nebular properties and the low-spatial resolution 
{\it IRAS} observations of NGC\,6888. We have shown that their SED 
overestimated the real IR flux from this WR nebula. 
Although the estimated mass is of the same order as that estimated by our
model (20~M$_{\odot}$), they used a blackbody model for the stellar radiation
field with $T_\mathrm{eff}=50$~kK. These authors restricted the large size of
their big grains to 0.25~$\mu$m but we found that increasing it to 0.5~$\mu$m
improved significantly our comparison with the IR SED.
The limit of 0.5~$\mu$m was set following the findings of 
\citet{Scicluna2015} on the RSG VY\,Canis Majoris. We note that a smaller
population of big grains did not
result in a good fit to the IR SED. 
In addition we tried models with dust sizes larger than 0.5~$\mu$m 
without significant improvement.
It is important to note that 93\% of the dust mass corresponds to 
grains with sizes similar as those obtained for RSG stars (larger
than those estimated for the ISM). 
That is, we are detecting the material ejected in the 
previous RSG phase with negligible contribution from swept 
up ISM material.

It seems indeed that the total mass of the nebula corresponds to material 
ejected by WR\,136 in its previous RSG phase without any contribution 
from swept-up ISM material as suggested by the analysis of the IR images. 
Adopting a current mass for WR\,136 of $\sim$22~M$_{\odot}$ 
\citep[see Table~5;][]{Hamann2019} one can estimate its initial mass to be 
$\lesssim$50~M$_{\odot}$. This is corroborated by the stellar evolutionary
models presented by \citet{Ek2012}\footnote{The models were retrived from 
the Geneva Code wavepage: 
\url{https://www.unige.ch/sciences/astro/evolution/en/database/}.}. 
In Figure~\ref{fig:tracks} we present HR
diagrams obtained by using stellar evolution models at $Z=0.014$ with 
and without rotation with initial masses between 32 and 
60~M$_{\odot}$. The stellar parameters ($T_\mathrm{eff}$ 
and $L$) estimated from the PoWR stellar atmosphere models
\citep[][]{Hamann2019} are somewhat consistent with WR\,136 having an 
initial mass of 50~M$_{\odot}$. 
Abundance estimates, in particular the He versus C/O
diagram presented by \citet{Esteban2016}, are also consistent with this finding 
(see Figure~9 in that paper). 
Finally, we calculated the averaged mass-loss rate predicted for a 
50~M$_{\odot}$ stellar evolution model. This resulted in $1.9\times10^{-4}$~M$_{\odot}$, very similar to that
estimated using our detailed model.

We note that other authors have estimated 
ionized masses of $\sim$4~M$_\odot$ 
\citep[see][]{Marston1988,Kwitter1981,Wendker1975} for NGC\,6888 
using radio and H$\alpha$ line emission. We have recalculated 
their estimates by following the procedure in \citet{Marston1988} 
using the H$\alpha$ intensity with the current distance of 
1.9~kpc and an electron density of 180~cm$^{-3}$ as obtained for 
the dominant shell in our model (also consistent with observations). 
This resulted in 21.9$^{+7.2}_{-5.0}$~M$_{\odot}$, which is 
consistent with the total estimated 
mass from our Cloudy model, but slightly smaller. 
This seems to suggest that NGC\,6888 is almost 
completely ionized. If the difference, 3.6~M$_{\odot}$, 
is considered to be neutral material, 
this should be shielded from the UV radiation from 
WR\,136 in the dense clumps and filaments seen in the H$\alpha +$[N\,{\sc ii}] images.

It is not necessary to invoke a contribution from the ISM
in the mass of NGC\,6888. An appropriate distance estimation by means of 
the {\it Gaia} observations, and improvement of the stellar evolution models,
and a careful dissection of the IR SED have allowed us to conclude that the
origin of this WR nebula is purely due to processed ejected 
stellar wind.

We would like to point out that in order to produce a consistent 
model of WR nebulae one must take into account important aspects such as 
an appropriate description of the WR stellar atmosphere of the progenitor
star and the simultaneous treatment of gas and dust (see Paper~I).
In particular we note here that 
the density distribution of our model was initially 
based on the findings reported by 
\cite{reyes2015} which only modeled the photoionized 
structure of NGC\,6888.
 
Although these authors used a very similar WR stellar 
atmosphere as that used in the present work 
(see Fig.~\ref{fig:sed_comparation}), the inclusion of 
dust in the model requires some tuning to the density parameters 
to simultaneously fit the optical and IR observations.

Finally, we note that our model predictions have 
been compared to single stellar evolution models,
but it is currently accepted that massive stars are 
born in binary systems affecting their evolution (see Section~\ref{sec:intro}). However, there is
no observational evidence that WR\,136 is a binary star \citep[see][]{Fullard2020,Grafener2012} and the clumpy 
morphology of its associated WR nebula seems to have
been the result of instabilities produced by the wind-wind
interaction scenario, in contrast to the ballistic 
expanding clumps expected in a common envelope 
stripping scenario (see discussion in Paper~I).

\section{Conclusions}

We presented a characterization of the distribution and properties 
of dust in the WR nebula NGC\,6888 around WR\,136. 
We used archival IR observations (spectra and images) that cover the 
3--160~$\mu$m wavelength range in conjunction with the photoionization 
code {\sc Cloudy} to model simultaneously the properties of the ionized 
and dust components in NGC\,6888. 
Our findings can be summarized as:
\begin{itemize}
    
\item The combination of the IR images allowed us 
to perform an unprecedented study of the spatial distribution of dust 
in NGC\,6888. The high-resolution IR images used here
helped us to dissect the contribution from dust in NGC\,6888 and that
corresponding to the ISM.
We discovered a dark region between the IR emission at the outer edge of the WR nebula 
coincident with the [O\,{\sc iii}] optical
emission. We suggest that this dark region is
unveiling the destruction of dust at the edge of NGC6888 
due to its relatively high expansion velocity ($\sim$80~km~s$^{-1}$). 

\item The IR photometry extracted taking into account the nebular morphology of
NGC\,6888 is heavily contaminated by the contribution from ISM clouds and 
filaments, a problem that could not be properly resolved in previous IR studies 
of this WR nebula. 
By studying the IR SED from different regions within NGC\,6888 and its 
immediate surroundings, we suggest that the NW region has no contamination 
from the ISM.
The true IR SED of NGC\,6888 peaks between 70~$\mu$m and 100~$\mu$m, very similar
to other WR nebulae.

\item Our best model to the nebular parameters and IR SED was achieved by adopting 
a spherical two-shell distribution with inner and outer densities of 
$n_1=400$~cm$^{-3}$ and $n_2=180$~cm$^{-3}$ and filling factors of 10$^{-3}$ and
0.07, respectively. We adopted two population of dust grains with 
sizes $a_\mathrm{small}$=[0.002--0.008]~$\mu$m and 
$a_\mathrm{big}$=[0.05--0.50]~$\mu$m, the later in line with the large dust grains
in the RSG star VY Canis Majoris. The total dust mass in NGC\,6888
resulted in 0.14$^{+0.03}_{-0.01}$~M$_\odot$ with a dust-to-gas ratio of $5.6\times 10^{-3}$. 
The small grains contribute to 6\% of the total dust mass.

\item We found that the dust in NGC\,6888 is dominated by large dust 
with RSG origin which suggest that this WR nebula is mainly composed 
by material ejected by the star with negligible contribution from 
swept up ISM.

\item The total estimated mass of NGC\,6888 is $25.5^{+4.7}_{-2.8}$~$M_{\odot}$, 
together with the estimates from \cite{Hamann2019} for its progenitor star, 
leads us to suggest an initial mass $\lesssim$50~$M_\odot$ 
for WR\,136. 
This result is supported by stellar evolution models, in particular the 
prediction of the mass-loss rate and the abundance determinations reported 
in the literature.

\end{itemize}

\section*{Acknowledgments} 

The authors would like to thank the referee, Anthony P. Marston, for comments and 
suggestions that improved the presentation of this paper. The authors are also thankful to
J.\,Reyes-P\'{e}rez for providing the {\sc cmfgen} model of WR\,136.
G.R. would like to thank S.J.\,Arthur and people at IRyA-UNAM for their support 
during the realization of this project. G.R., E.S and J.A.Q.-M. 
acknowledge support from Consejo Nacional de Ciencia y Tecnolog\'{i}a (CONACyT)
for student scholarship. J.A.T., G.R., and M.A.G. are funded by
UNAM DGAPA PAPIIT project IA100720.
M.A.G. acknowledges support of the
Spanish Ministerio de Ciencia, Innovación y Universidades grant
PGC2018-102184-B-I00, co-funded by FEDER funds. G.R.-L. acknowledges 
support from CONACyT and PRODEP (Mexico).
VMAGG acknowledges support from the Programa de Becas 
posdoctorales of DGAPA UNAM.
This work makes use of {\sc iraf}, 
distributed by the National Optical Astronomy Observatory, which is 
operated by the Association of Universities for Research in Astronomy under
cooperative agreement with the National Science Foundation. This work makes use
of {\it Herschel}, {\it Spitzer} and {\it WISE} IR observations. 
{\it Herschel} is an ESA space observatory 
with science instruments provided by European-led Principal Investigator 
consortia and with important participation from NASA. The
{\it Spitzer} Space Telescope was operated by the 
Jet Propulsion Laboratory, California Institute of Technology 
under a contract with NASA. Support for this work was provided 
by NASA through an award issued by JPL/Caltech. {\it WISE} is a joint 
project of the University of California (Los Angeles, USA) 
and the JPL/Caltech, funded by NASA.

\section*{Data availability}

 The data underlying this article will be shared on reasonable 
request to the corresponding author.

{}

\appendix

\section{Dust destruction at the edge of NGC\,6888}
\label{sec:appA}

\begin{figure}
\begin{center}
\includegraphics[width=\linewidth]{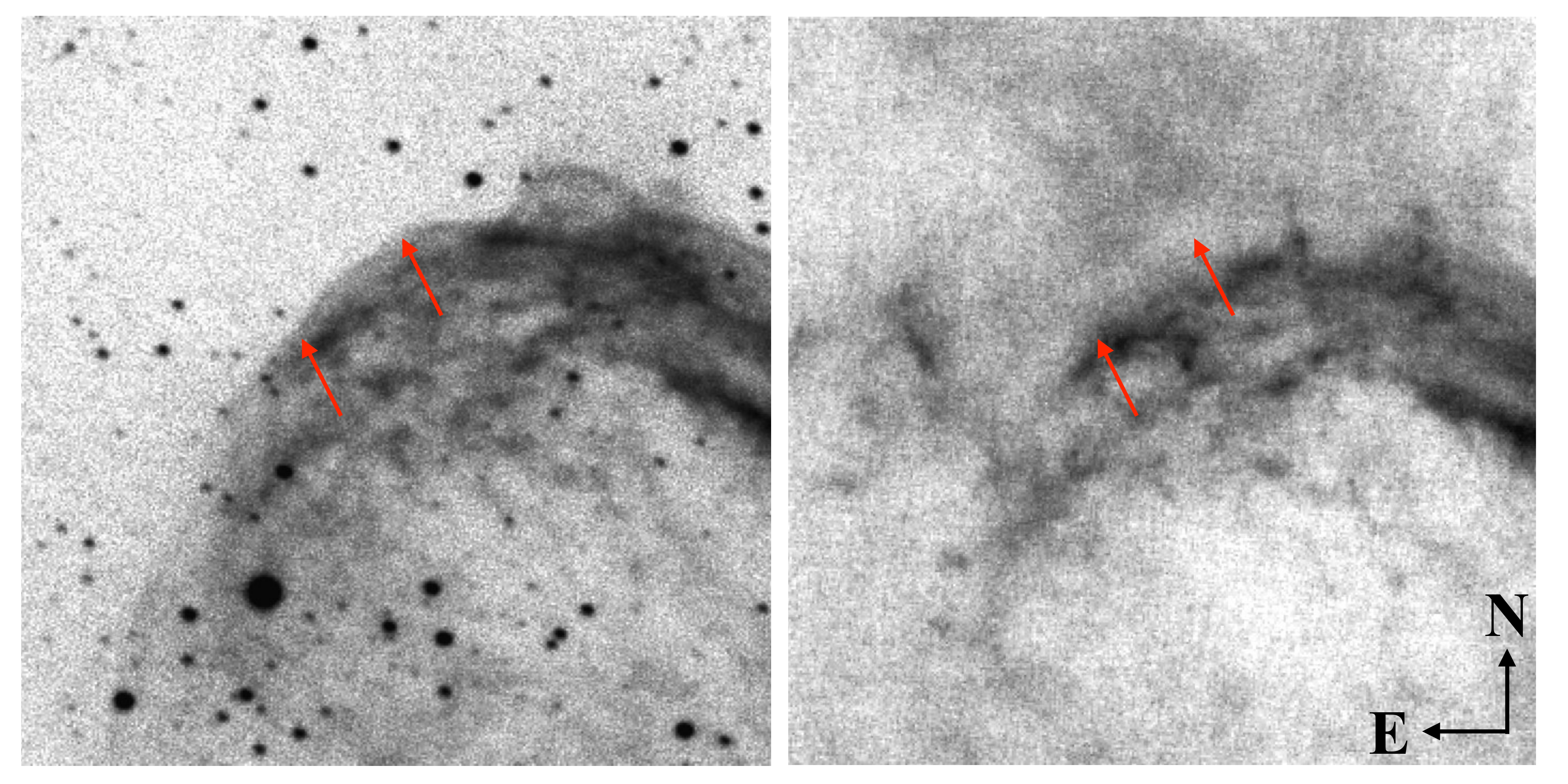}
\caption{Close-up of the NE cap of NGC\,6888. 
Grey scale images of the optical [O\,{\sc iii}] narrow-band filter (left) and 
the {\it Herschel} PACS 70~$\mu$m (right) showing the gap (lack of emission) described 
in Section~3. The red arrows show the edge of the [O\,{\sc iii}] in both panels. Both
panels have the same FoV.}
\label{fig:fig_gap}
\end{center}
\end{figure}

To further illustrate the correlation between the gap region surrounding
NGC\,6888 detected in IR images and that of the nebular emission, 
we have created close-up, grey-scale images of the optical [O\,{\sc iii}] and that
of the {\it Herschel} PACS 70~$\mu$m of the NE region. 
The images presented in Figure~\ref{fig:fig_gap}, show with red arrows the position
of the [O\,{\sc iii}] emission.

\end{document}